\documentclass[lettersize,journal]{IEEEtran}
\usepackage{amsmath,amsfonts}
\usepackage{algorithmic}
\usepackage{algorithm}
\usepackage{array}
\usepackage[caption=false,font=normalsize,labelfont=sf,textfont=sf]{subfig}
\usepackage{textcomp}
\usepackage{stfloats}
\usepackage{url}
\usepackage{verbatim}
\usepackage{graphicx}
\usepackage{cite}
\usepackage{siunitx}
\usepackage{amssymb}
\usepackage{xcolor}
\usepackage[acronym]{glossaries}
\usepackage{hyperref}
\usepackage{rotating}
\usepackage{multirow}
\usepackage{arydshln}
\usepackage{orcidlink}
\usepackage{booktabs}

\DeclareMathOperator{\LN}{LN}

\DeclareMathOperator{\conv2d}{Conv2D}

\DeclareMathOperator{\sigmoid}{sigmoid}
\DeclareMathOperator{\FFT2D}{FFT2D}

\DeclareMathOperator{\pixelshuffle}{PixelShuffle}
\DeclareMathOperator{\rstb}{RSTB}
\DeclareMathOperator{\stl}{STL}

\DeclareMathOperator{\comp}{comp}

\newacronym{DLSS}{DLSS}{deep learning super sampling}
\newacronym{FFW}{FFW}{feed-forward}
\newacronym{PRI}{PRI}{Pulse Repetition Interval}
\newacronym{GPU}{GPU}{Graphics Processing Unit}
\newacronym{CW}{CW}{continuous wave}
\newacronym{TOA}{TOA}{time of arrival}
\newacronym{TDOA}{TDOA}{time difference of arrival}
\newacronym{AOA}{AOA}{angle of arrival}
\newacronym{RF}{RF}{radio frequency}
\newacronym{PW}{PW}{pulse width}
\newacronym{BW}{BW}{bandwidth}
\newacronym{PA}{PA}{pulse amplitude}
\newacronym[plural = RNNs]{RNN}{RNN}{recurrent neural network}
\newacronym{SNR}{SNR}{signal-to-noise ratio}
\newacronym[plural = STFTs]{STFT}{STFT}{short-time Fourier transform}
\newacronym{FFT}{FFT}{fast Fourier transform}
\newacronym{iFFT}{iFFT}{inverse fast Fourier transform}
\newacronym{DP-TF-transformer}{DP-TF-transformer}{dual-path time-frequency transformer}
\newacronym[plural = iSTFTs]{iSTFT}{iSTFT}{inverse short-time Fourier transform}
\newacronym{LN}{LN}{layer normalization}
\newacronym{ReLU}{ReLU}{rectified linear unit}
\newacronym{FMCW}{FMCW}{frequency-modulated continuous-wave}
\newacronym{CPI}{CPI}{coherent processing interval}
\newacronym{CS}{CS}{chirp-sequence}
\newacronym{CFAR}{CFAR}{constant false-alarm rate}
\newacronym{CA-CFAR}{CA-CFAR}{cell-averaging constant false-alarm rate}
\newacronym{OS-CFAR}{OS-CFAR}{ordered statistic constant false-alarm rate}
\newacronym{GAN}{GAN}{generative adversarial network}
\newacronym{rD}{rD}{range-Doppler}
\newacronym{rDa}{rDa}{range-Doppler-angle}
\newacronym{RCS}{RCS}{radar cross section}
\newacronym{ViT}{ViT}{vision transformer}
\newacronym{SWIN}{SWIN}{shifted window}
\newacronym{DPSWIN}{DPSWIN}{dual-path shifted window}
\newacronym{MHA}{MHA}{multi-head attention}
\newacronym{SSM}{SSM}{state-space model}
\newacronym{ADC}{ADC}{analog-to-digital converter}
\newacronym{MSE}{MSE}{mean squared error}
\newacronym{RMSE}{RMSE}{root mean squared error}
\newacronym{FA}{FA}{false-alarm}
\newacronym{LiDAR}{LiDAR}{light detection and ranging}
\newacronym{STL}{STL}{Swin transformer layer}
\newacronym{RSTB}{RSTB}{residual Swin transformer block}
\newacronym{cGAN}{cGAN}{conditional generative adversarial network}
\newacronym{CUT}{CUT}{cell under test}
\newacronym{BCE}{BCE}{binary cross entropy}
\newacronym{KLD}{KLD}{Kullback-Leibler divergence}
\newacronym{PRT}{PRT}{pulse repetition time}
\newacronym{SAR}{SAR}{synthetic aperture radar}
\newacronym{CNN}{CNN}{convolutional neural network}
\newacronym{PPI}{PPI}{plan position indicator}
\newacronym{NEXRAD}{NEXRAD}{next generation weather radar}
\newacronym{LSTM}{LSTM}{long short-term memory}
\newacronym{MUSIC}{MUSIC}{multiple signal classification}
\newacronym{RELAX}{RELAX}{relaxation}
\newacronym{CVD}{CVD}{cadence velocity diagram}
\newacronym{RMT}{CVD}{cadence velocity diagram}
\newacronym{UAV}{UAV}{unmanned aerial vehicle}
\newacronym{DP}{DP}{dual-path}
\newacronym{MSA}{MSA}{multi-head self-attention}
\newacronym{MLP}{MLP}{multi-layer perceptron}
\newacronym{GELU}{GELU}{Gaussian error linear unit}
\newacronym{Adam}{Adam}{adaptive moment estimation}
\newacronym{IoU}{IoU}{intersection over union}
\newacronym{LSD}{LSD}{log-spectral distance}
\newacronym{MAE}{MAE}{mean absolute error}
\newacronym{ResNet}{ResNet}{residual network}
\newacronym{DETR}{DETR}{detection transformer}
\newacronym{ECG}{ECG}{electrocardiogram}
\newacronym{DoA}{DoA}{direction of arrival}
\newacronym{CRB}{CRB}{Cramér–Rao bound}
\newacronym{SOTA}{SOTA}{state-of-the-art}
\glsaddall

\begin{document}

\title{Super-Resolution of Range-Doppler Maps: \\A Case Study with Chirp-Sequence Radar and Transformer}
\author{Sven Hinderer\textsuperscript{\orcidlink{0009-0008-5230-3034}}, Jonathan Riese\textsuperscript{\orcidlink{0009-0001-0905-6129}}, Zheming Yin\textsuperscript{\orcidlink{0009-0008-3291-0523}}, Bin Yang\textsuperscript{\orcidlink{0000-0002-8322-117X}},~\IEEEmembership{Senior Member,~IEEE}

\thanks{This work was conducted at the Institute of Signal Processing and System Theory, University of Stuttgart, Germany. The corresponding author can be contacted via email at
{\tt sven.hinderer@iss.uni-stuttgart.de}.}
%\thanks{Manuscript received Month XX, XXXX; revised Month XX, XXXX.}
}

%\markboth{IEEE Transactions on Radar Systems,~Vol.~XX, No.~X, August~XXXX}%

%\IEEEpubid{\color{red}0000--0000/00\$00.00~\copyright~2021 IEEE\color{black}}

\maketitle

\begin{abstract}
\Gls{rD} maps produced by chirp-sequence (CS) radar systems are fundamentally limited in resolution by bandwidth, carrier frequency, and coherent processing interval constraints. Improving resolution through hardware is often impractical due to regulatory, cost, and real-time operation requirements.

In this work, we investigate deep learning-based super-resolution of rD maps in both range and Doppler using a real-world dataset collected with an Infineon millimeter-wave CS radar. We propose a memory-efficient transformer architecture based on \gls{DPSWIN} attention, which combines axial/dual-path attention with 1D \gls{SWIN} attention for scalable processing of high-dimensional radar data. Our model is benchmarked against existing 2D SWIN attention-based super-resolution models, which we adapt to the rD-map super-resolution task.

We prioritize reconstruction fidelity over perceptual quality and employ \gls{RMSE}-based training objectives. We avoid adversarial or perceptual losses that may introduce visually plausible but physically incorrect structures. In addition, we incorporate CFAR-based target-detection losses to optimize downstream target detectability in the super-resolved rD maps.

We further study the impact of signal processing and training design choices on the super-resolution, including magnitude compression, spatial upsampling strategies, and loss formulations. Experimental results demonstrate that the proposed framework achieves computationally efficient rD map super-resolution on previously unseen real-world environments.	
\end{abstract}


\section{Introduction}
High-resolution radar sensing is essential for accurate parameter estimation, but hardware constraints such as limited bandwidth often restrict achievable range and Doppler resolution. Classical super-resolution methods, including \gls{MUSIC} \cite{music} and related model-based approaches, can improve spectral resolution but rely on strong assumptions about signal statistics, target structure, and noise characteristics. In complex real-world environments such as indoor spaces, these assumptions are frequently violated. Moreover, the computational complexity of such methods limits their real-time applicability for high-resolution \gls{rD} maps.

Deep learning offers an alternative by learning the super-resolution mapping directly from data and enabling efficient inference through a single forward pass. While super-resolution has been extensively studied in computer vision \cite{image_sr_review}, radar rD map super-resolution remains comparatively underexplored. Existing radar works mainly focus on simplified scenarios, including simulated point targets or controlled reflector measurements, and predominantly rely on convolutional neural networks (CNNs) \cite{rD_diffusion, rD_superres_unet, phd_chenming, rD_upscale_complex}. 

In contrast, we investigate rD map super-resolution using real radar measurements collected in challenging indoor and outdoor pedestrian environments and study transformer-based architectures tailored to radar signal characteristics.

Our primary objectives are accurate reconstruction of signal power and correct target detection in super-resolved rD maps. In radar applications, false targets can severely degrade downstream tasks such as detection and tracking, particularly in safety-critical settings. For this reason, we avoid adversarial or perceptual losses that improve perceptual quality. Perceptual losses computed with feature-maps extracted from pre-trained models can introduce artifacts. Adversarial losses that are used in \glspl{GAN} are known to increase hallucination risk and signal distortion 
\cite{perc_dist_tradeoff, dmatching_hallu, SRGAN, pix2pix}, making them less suitable for our application. Our primary loss function \eqref{eq:trainrmse} is based on the \gls{RMSE} between the super-resolved rD bin magnitudes and the corresponding high-resolution ground truth to minimize the distortion in the reconstructed rD maps. 

We investigate how domain-specific signal-processing choices influence super-resolution performance. In particular, we study:

\begin{itemize}
	\item integration of CFAR-based losses to improve downstream target detection and reduce blurring,
	\item the effect of rD bin magnitude compression strategies (e.g. linear scale vs. log-scale)  for handling the large dynamic range of rD maps,
	\item and signal-processing choices such as zero-padding before \gls{FFT} computation.
\end{itemize}

Most radar research, including the related works discussed in Sec.~\ref{sec:rel_works}, focuses on either linear-scale or log-scale signal representations, with \cite{rD_diffusion} being an exception by testing both. The linear-scale approaches are often not compared against log-scale alternatives, despite the latter being better suited for handling signals with large dynamic ranges. Moreover, compression functions beyond the base-ten logarithm remain largely unexplored in radar signal processing. In this work, we investigate this overlooked pre-processing step and demonstrate that simple adaptations on the compression function can yield additional performance gains.

We further propose an efficient transformer attention mechanism specifically designed for rD map processing. Our method combines dual-path attention along range and Doppler dimensions with shifted-window attention to achieve scalable computation while preserving spectral structure.

The main contributions of this work are:

\begin{itemize}
	\item Real-world radar dataset:
	We collect a real radar dataset using an Infineon chirp-sequence radar in indoor and outdoor pedestrian environments and generate paired low- and high-resolution rD maps through controlled baseband sub-sampling \cite{phd_chenming}.
	\item Efficient transformer architecture for radar super-resolution:
	We propose \gls{DPSWIN} attention, which combines dual-path attention \cite{axial_attention, dptnet, SepFormer} with 1D \gls{SWIN} attention \cite{swin, swin_1D_time_series, swin_1D_hrr}. Compared to conventional 2D SWIN transformers, DPSWIN significantly reduces computational and memory complexity while preserving performance on rD map super-resolution.
	\item CFAR-guided super-resolution training:
	We improve CFAR-based loss functions \cite{cfar_loss} and extend them to the rD map super-resolution task. We demonstrate that integrating target-detection objectives into training improves rD map quality.
	\item Spatial upsampling strategies: We show that \gls{FFT} zero-padding \cite{az_superres, rD_diffusion} outperforms pixel-shuffle upsampling \cite{pixel_shuffle} for our task.
	\item Compression: We benchmark using linear-scale, log-scale, and power-law compressed rD maps and show that power-law compression gives overall best results.
\end{itemize}

In summary, our results demonstrate that accurate and computationally efficient super-resolution of radar rD maps is feasible even in challenging real-world environments.

The article is structured as follows: Sec.~\ref{sec:rel_works} reviews the state of the art on deep learning-based radar super-resolution and transformers for radar. Sec.~\ref{sec:cs_radar} gives an introduction to \gls{CS} radar and shows the data generation process for supervised rD super-resolution. It also contains a precise task description. Sec.~\ref{sec:sp_chain} describes the pipeline and architecture used for rD super-resolution. Sec.~\ref{sec:cfar} contains a short overview of CFAR target detection and derives the CFAR losses. Sec.~\ref{sec:exp} describes the hardware, dataset, training and evaluation. Sec.~\ref{sec:res} shows our results. Sec.~\ref{sec:con} summarizes the article including current limitations and gives an outlook for future work.

\section{Related works}
\label{sec:rel_works}
Although various deep learning approaches exist that upsample radar signals, the body of work is still relatively limited, especially compared with larger and more deep learning-focused communities such as computer vision or audio. Radar super-resolution has the additional difficulty that it is a very broad subject since the input data varies greatly based on task and sensor. Upsampling \gls{rD} maps is e.g. different from upsampling \gls{SAR} images, or point-clouds etc., even though the underlying idea is identical. 

We only consider related literature that processes raw time-domain data, its spectra or radar images, as those works are closest to ours. We omit the point-cloud literature, which often works on sparse and pre-processed data to handle the large data volume. While popular network architectures from related tasks in other domains generally also work well for radar, radar signals have special properties including large dynamic range of signal power, sparsity, background noise, or speckle noise, that require special handling.

\subsection{CNNs and U-Nets for radar super-resolution} 
\Glspl{CNN}, especially the well known convolutional U-Net \cite{unet} and its derivatives, are found most in other radar super-resolution works. A \gls{ResNet} has been applied as generator for \gls{rD} super-resolution in a \gls{cGAN} framework in \cite{phd_chenming} and evaluated on simulated point targets. We adopt the (low-res, high-res) tuple generation process from this work, but we use real data. In \cite{rD_superres_unet}, a similar approach with a U-Net generator and a convolutional discriminator is found, upsampling Doppler in \gls{rD} maps. However, this also contains only simulated data and simplified real data with corner reflectors as point targets. A serial stack of U-Nets is used in the generator in \cite{radar_superres_karim_sherif} for spectrogram and azimuth super-resolution. They also employ a \gls{cGAN} for higher realism of the super-resolution data and derive a perceptual loss with the discriminator's convolutional feature maps. 

\cite{azimuth_superres_rDa} employs a U-Net trained on range-azimuth features extracted from low-resolution \gls{rDa} maps to reconstruct corresponding high-resolution range-azimuth maps. \cite{az_superres} proposes two networks capable of azimuth super-resolution. One \gls{CNN} that works directly on the time-domain baseband samples, while another U-Net works on the linear-scale range-azimuth maps. A complex-valued \gls{CNN} upsamples range in multi-band 3D \gls{SAR} in \cite{sar_superres_cvcnn}, where stacked blocks including \glspl{FFT} and \glspl{iFFT} process linear-scale time- and frequency-domain baseband data sequentially. Other complex-valued \glspl{CNN} with raw complex-valued baseband inputs are used in \cite{cvnn_hrr} for super-resolution of simulated high-resolution range profiles, and in \cite{rD_upscale_complex} for simulated \gls{rD} maps.

Further U-Net applications for radar super-resolution are \gls{NEXRAD} \gls{PPI} scans in \cite{ppi_unet}, or cross-modal radar-to-\gls{LiDAR} upsampling in \cite{4d_superres}. 

A diffusion model \cite{diffusion} with a U-Net has recently been proposed to upsample Doppler in \gls{rD} maps \cite{rD_diffusion}. While diffusion models achieve \gls{SOTA} performance in many offline tasks and they might outperform our model, they are less suitable for real-time processing due to their iterative inference process.

We initially experimented with U-Nets and simple CNNs, but preliminary experiments have shown that they consistently perform worse than our transformer networks with comparable parameter count. Transformers outperforming U-Nets is supported by automotive radar research \cite{trans_radar, radar_foundation}.  Due to their worse performance, we did not consider U-Nets for further investigation.

\subsection{Transformers for radar}
\label{sec:lit_transformer_radar} 
Transformers \cite{transformer} have been the dominating deep learning architecture in the past years. Transformers combine highly parallelized training and inference with small inductive bias and great scaling with large data. Their greatest weakness is their quadratic complexity with regard to the input length due to the attention mechanism. This complexity can be reduced, e.g. by applying attention locally not globally.

In radar, transformers have previously been used mostly for tasks other than super-resolution.
A 3D \gls{CNN}, 3D \gls{SWIN} transformer combination for object detection is described in  \cite{radar_swin_3D_object_detection}. Our model will also use such shifted attention windows, but in an efficient 1D axial version described in Sec.~\ref{sec:swin}. Other \gls{SWIN}-based models are used e.g. for object detection in \cite{radar_swin_od} and for target recognition in high-resolution range profiles in \cite{radar_swin_hrr}. Another object detection transformer is introduced in \cite{radar_trans_od_vit_prior}, where they applied a \gls{ViT} \cite{vit}  extension with spatial prior to better reflect radar signal characteristics. In \cite{vit_radar_gait_recognition}, \glspl{ViT} are applied to spectrograms and \glspl{CVD} for gait recognition. There's further work for radar indoor perception in \cite{radar_trans_indoor} with a \gls{DETR}-based network \cite{detr}, including various radar- and problem-specific modifications.

Motivated by some research works on audio signals, we previously applied an axial/\gls{DP} transformer \cite{axial_attention,SepFormer,dptnet} to radar signal separation \cite{irs22_radar_separation}. This model partitions a 2D input into non-overlapping 1D axial windows, applies attention independently within each window, and achieves global attention by performing this operation along both axes (see Sec.~\ref{sec:dp} for details). We further reduce the complexity of dual-path attention with a more efficient 1D axial shifted window attention mechanism named \gls{DPSWIN} in this work.
In \cite{trans_radar}, a model was proposed for semantic segmentation that also applies dual-path attention. They extended this idea by proposing adaptive, directional attention with a learnable aggregation operation before the attention that can span over multiple rows/columns. Their evaluation shows that their model outperforms the 3D SWIN in \cite{radar_swin_3D_object_detection} in most metrics at much smaller model size. Another model similar to dual-path transformers was used in \cite{dp_radar_people_counting} for people counting, where individual transformers are applied separately to 1D time- and frequency-domain baseband signals. 

\subsection{Transformers for radar super-resolution}
The only works using transformers in radar super-resolution are in \gls{SAR} imaging. The work in \cite{swin_1D_hrr} upsampled the range profile of SAR images using 1D SWIN. We also apply 1D SWIN, but for both the range axis and the Doppler axis. However, their model was trained on simulated data only. They also train on complex-valued baseband signals, which we believe is suboptimal for most radar applications due to the large dynamic range (see Sec.~\ref{sec:compression}). Other works applied to full \gls{SAR} images use a \gls{cGAN} with a transformer generator \cite{superres_sar}, a \gls{ViT} \cite{sar_super_vit}, or a \gls{SWIN} transformer \cite{radar_super_sar2}. The evaluation in \cite{superres_sar} includes a SwinIR \cite{SwinIR} model, originally applied to images, which shows good performance. Our model is inspired by SwinIR, but has multiple modifications explained in Sec.~\ref{sec:sp_chain}. We believe we are the first to apply transformers for \gls{rD} super-resolution.

\section{Chirp-sequence radar and data generation}
\label{sec:cs_radar}

\gls{CS} radars like our Infineon BGT60TR13C \cite{infineon2023bgt60tr13c_datasheet} are \gls{FMCW} radars with steep frequency ramps, allowing for low-cost range, Doppler (and possibly also angle and polarimetry) estimation. In addition, \gls{CS} radar offers high range resolution and robustness to narrowband interference by using large bandwidth and high \gls{SNR} at low peak transmit power by transmitting continuous waves (as opposed to pulses) and (coherently) integrating multiple received chirps. At millimeter wavelength, a \gls{CS} radar also becomes very compact due to the small antenna size. The BGT60TR13C works in the 60 GHz band, i.e. the wavelength is $\lambda = c/f_c \approx 5\,$mm, where $c$ is the speed of light and $f_c= 61\,$GHz is the center frequency of our radar signals. This sensor and similar radar sensors are also studied for different tasks including tracking \cite{bgt_rl_tracking}, gesture recognition \cite{bgt_gesture_recogn, bgt_gesture_recog2}, activity recognition \cite{bgt_activ_recog, bgt_activ_class}, and indoor localization \cite{bgt_indoor_loc, irs25}.

\gls{CS} radars transmit a continuous train of (usually) linear chirps. Each chirp has a duration $T_{chirp}$, bandwidth $B$, and center frequency $f_c$. The time between chirps is the \gls{PRT} $T_{rep}$, with the naming convention coming from pulse radar. The transmitted signal is reflected by the environment, leading to time-delayed versions of the transmitted chirps at the receiver. The transmitted and received signals are mixed, filtered, amplified, and digitized, resulting in the high-resolution baseband signal of a 2D time-domain frame $\mathbf{Y}^{hr}_t\in\mathbb{R}^{(2N_r)\times N_D}$ collected at time $t$. The frame duration is called \gls{CPI} $\mathrm{CPI}=T_{rep}N_D$. $N_r$ is the number of range bins and $N_D$ is the number of velocity/Doppler bins of the corresponding high-resolution rD spectrum $\mathbf{X}^{hr}_t=|\FFT2D(\mathbf{Y}^{hr}_t)|\in\mathbb{R}^{(N_r+1)\times N_D}$. The change in the number of range bins from $2N_r$ to $N_r+1$ comes from the omission of Hermitian symmetric negative range bins, as the BGT60TR13C uses a real-valued baseband signal (only one In-phase channel) and we apply the range FFT to it. The additional range bin is later dropped to get the range dimension $N_r$. The Doppler FFT is applied to the complex-valued range FFT outputs, thus no change in Doppler dimension occurs. 

\begin{figure}
	\centering
	\includegraphics[width=0.95\linewidth]{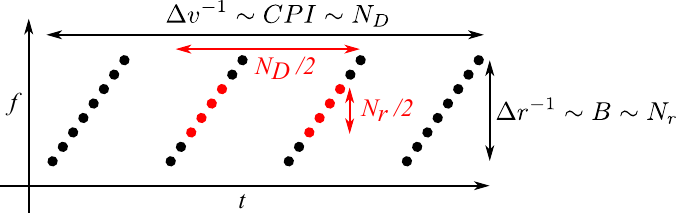}
	\caption{Depiction of the (low-res, high-res) tuple generation. The high-resolution chirp sequence (black) is downsampled by factor two in both dimensions to create the low-resolution chirp sequence (blue). Dropping half of the slow-time samples (chirps - time axis) reduces the velocity resolution by factor two and dropping half of the fast-time samples (frequency axis) reduces the range resolution by factor two.}
	\label{fig:downsampling}
\end{figure}

The range resolution $\Delta r$ and velocity resolution $\Delta v$ are
\begin{equation}
	\Delta r = c / (2B),\,\,\Delta v = \lambda / (2 T_{rep}N_D).
\end{equation}
For the dataset generation of our supervised super-resolution task, we record the ground truth high-resolution baseband signal. The corresponding low-resolution signal can then be easily created by dropping consecutive fast-time samples to reduce range resolution and by dropping slow-time samples to reduce velocity resolution. Our (low-res, high-res) tuple generation process is depicted in Fig.~\ref{fig:downsampling} using the transmitted chirp-sequence. The respective components are dropped in the received 2D time-domain frame. In this work, we restrict us to a downsampling factor of two in both dimensions.
Our process varies slightly from the process proposed in \cite{phd_chenming}, which took the first half of all chirps. We shifted this to the middle to reduce the maximum time deviation between low- and high-resolution signals. 

The task of the super-resolution network is then to recover those dropped fast- and slow-time samples, which we will do in frequency-domain, not in time-domain, i.e. in the \gls{rD} spectrum $\mathbf{X}$ after range and Doppler \gls{FFT}. 

\section{Range-Doppler super-resolution processing}
\label{sec:sp_chain}
\begin{figure}
	\centering
	\includegraphics[width=0.95\linewidth]{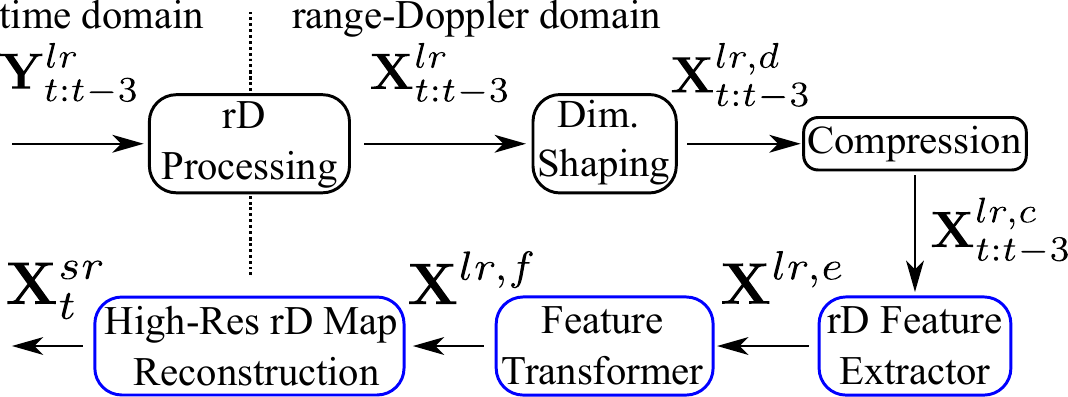}
	\caption{Overview of our signal processing chain consisting of pre-processing (black boxes) and trainable modules (blue boxes).}
	\label{fig:overview}
\end{figure}

An overview of our signal processing chain is given in Fig.~\ref{fig:overview}. Below, we first give a short functional explanation of the modules in Fig.~\ref{fig:overview}. Each module will then be described in detail.

The input $\mathbf{Y}^{lr}_{t:t-3}
\in
\mathbb{R}^{N_r \times \frac{N_D}{2} \times 4}$ contains a stack of the most recent low-resolution, time domain frame $\mathbf{Y}^{lr}_{t}
\in
\mathbb{R}^{N_r \times \frac{N_D}{2}}$ and three preceding ones. A low-resolution frame $\mathbf{Y}^{lr}_{t}$ is created by downsampling the corresponding high-resolution frame $\mathbf{Y}^{hr}_t$ that is our ground truth. Consecutive frames are recorded with a delay of $\SI{100}{ms}>\mathrm{CPI}$ to leave time for post-processing between frames. The dynamic scene changes across frames, where targets that are unresolved in the current frame $\mathbf{Y}^{lr}_{t}$ might have been resolvable in previous ones. The network can therefore exploit such temporal cues to improve rD super-resolution. Using $\mathbf{Y}^{lr}_{t:t-3}$ instead of $\mathbf{Y}^{lr}_{t}$ as input gave better results. $\mathbf{Y}^{lr}_{t:t-3}$ is therefore our input for all experiments. We omit a detailed investigation of this temporal aspect in this work.

The processing chain outputs the (compressed) super-resolved rD bin magnitudes $\mathbf{X}^{sr}_{t}\in\mathbb{R}^{N_r\times N_D}$ corresponding to the most recent low-resolution frame $\mathbf{Y}^{lr}_{t}$ and the high-resolution frame $\mathbf{Y}^{hr}_{t}$. Our goal is for $\mathbf{X}^{sr}_t$ to closely approximate the (compressed) high-resolution rD bin magnitudes $\mathbf{X}^{hr}_{t}$. In this study, we focus on target detection in rD maps and restrict the rD map processing to rD bin magnitude. An extension with phase information as e.g. required for successive \gls{DoA} estimation is possible, but out of the scope of this work.

Following the radar code examples provided by Infineon, we remove the bias in the fast-time samples of each individual chirp. Each frame in $\mathbf{Y}^{lr}_{t:t-3}$ is then transformed into a rD map by range-Doppler processing, giving the low-resolution rD map stack $\mathbf{X}^{lr}_{t:t-3}\in\mathbb{R}^{(N_r+1)\times N_D\times 4}$ with $\mathbf{X}^{lr}_{t}=|\FFT2D(\mathbf{Y}^{lr}_{t})|\in\mathbb{R}^{(N_r+1)\times N_D}$. If zero-padding is applied before the range/Doppler FFTs of the low-resolution signal, the shape of $\mathbf{X}^{lr}_t$ matches the shape of the corresponding high-resolution rD map $\mathbf{X}^{hr}_t\in\mathbb{R}^{(N_r+1)\times N_D}$. The exact processing and the case without zero-padding are explained below in Sec.~\ref{sec:rD_proc}.

We then drop the Nyquist frequency bin along range in the dimension shaping module (Sec.~\ref{sec:dim_shaping}) to convert the range dimension from $(N_r+1)$ to $N_r$. This gives  $\mathbf{X}^{lr,d}_{t:t-3}\in\mathbb{R}^{N_r\times N_D\times 4}$ and $\mathbf{X}^{hr}_t\in\mathbb{R}^{N_r\times N_D}$. Next, all rD bin magnitudes in $\mathbf{X}^{lr,d}_{t:t-3}$ are compressed as described in Sec.~\ref{sec:compression}. 

This pre-processed temporal rD map stack $\mathbf{X}^{lr,c}_{t:t-3}\in\mathbb{R}^{N_r\times N_D\times 4}$ is then fed to a convolutional feature extractor to expand the channel dimension, giving $\mathbf{X}^{lr,e}\in\mathbb{R}^{N_r\times N_D\times C}$  (Sec.~\ref{sec:feat_extr}). $\mathbf{X}^{lr,e}$ is processed by the feature transformer (Sec.~\ref{sec:feat_transform}) to transform the features into an improved representation for the following super-resolution step. The transformed features $\mathbf{X}^{lr,f}\in\mathbb{R}^{N_r\times N_D\times C}$ are then mapped to the compressed super-resolution rD map $\mathbf{X}^{sr}_t\in\mathbb{R}^{N_r\times N_D}$, which approximates the ground truth high-resolution rD map $\mathbf{X}^{hr}_t$, by the high-resolution range-Doppler map reconstruction block (Sec.~\ref{sec:extension}).

Our trainable blocks are inspired by SwinIR \cite{SwinIR}. SwinIR is a transformer-based architecture developed for image restoration tasks like super-resolution and is based on SWIN attention \cite{swin}. SWIN attention is explained in Sec.~\ref{sec:feat_transform}. We chose the original SwinIR and SWIN attention for our network basis. There have been architectural improvements, e.g. by changing the attention mechanism \cite{swinv2}, or by integrating spectral layers \cite{swinfir} to better capture global context. There also exist improved windowing mechanisms with neighborhood attention \cite{neighborhood_attn}. We restrict our network architecture study to the impact of the windowing mechanism. We investigate the use of low-complexity 1D instead of 2D shifted windows to exploit spectral rD map structure.

\subsection{Range-Doppler processing}
\label{sec:rD_proc}
We start with the rD processing and dimension shaping blocks in Fig.~\ref{fig:overview}. We have two versions, with and without zero-padding before the respective range and Doppler FFT, to study which one is better for super-resolution. With zero-padding, the $N_r$ fast-time samples of each chirp in a frame $\mathbf{Y}^{lr}_{t}\in\mathbb{R}^{N_r\times \frac{N_D}{2}}$ are first multiplied with a specified range window and then zero-padded from length $N_r$ to $2N_r$. This leaves $N_r+1$ non-negative range bins after the range FFT. Similarly, we apply windowing and zero-padding from $N_D/2$ to $N_D$ slow-time samples along the Doppler axis before the Doppler FFT, leaving $N_D$ Doppler bins after the FFT, hence $\mathbf{X}^{lr}_{t:t-3}\in\mathbb{R}^{(N_r+1)\times N_D\times 4}$.

Zero-padding before an FFT neither adds any information nor reduces the main lobe width of targets, but produces the wanted super-resolution rD map dimension by oversampling the rD spectrum. The reduced \gls{FFT} bin width can reduce the frequency offset between true target peaks and the closest FFT bin \cite{richards_modern_radar}. This frequency offset reduction, and the availability of more samples e.g. for peak interpolation, can improve the accuracy of peak frequency and magnitude estimation. Similarly, zero-padding may help in our problem of estimating super-resolved rD bin magnitudes. 

\begin{figure*}[h!]
	\centering
	\subfloat[Low-res: w.o. zero-padding.]{%
		\includegraphics[width=0.3\textwidth]{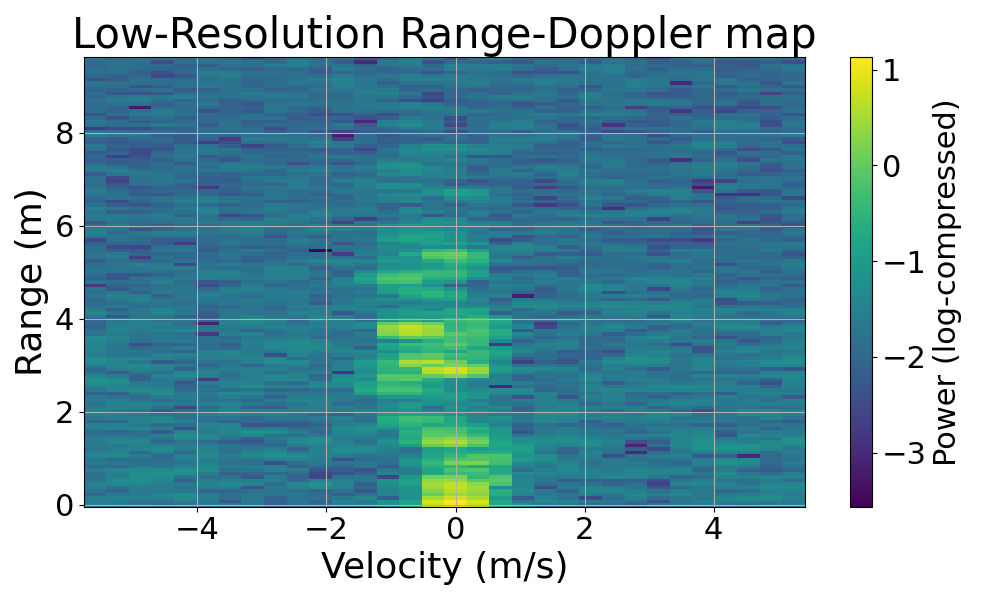}%
	}\hspace{0.03\textwidth}%
	\subfloat[Low-res: with zero-padding.]{%
		\includegraphics[width=0.3\textwidth]{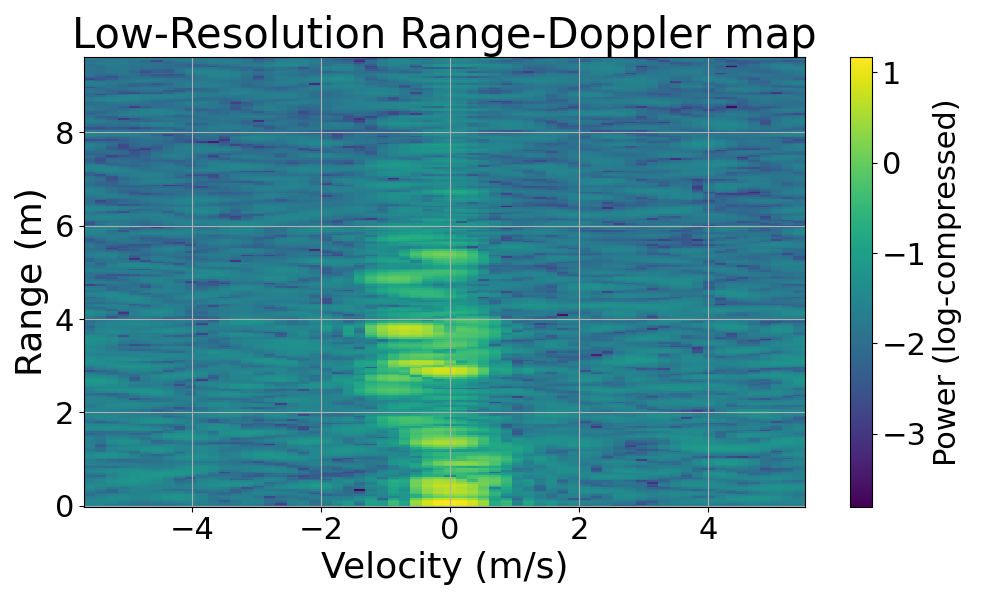}%
	}\hspace{0.03\textwidth}%
	\subfloat[High-res: w.o. zero-padding.]{%
		\includegraphics[width=0.3\textwidth]{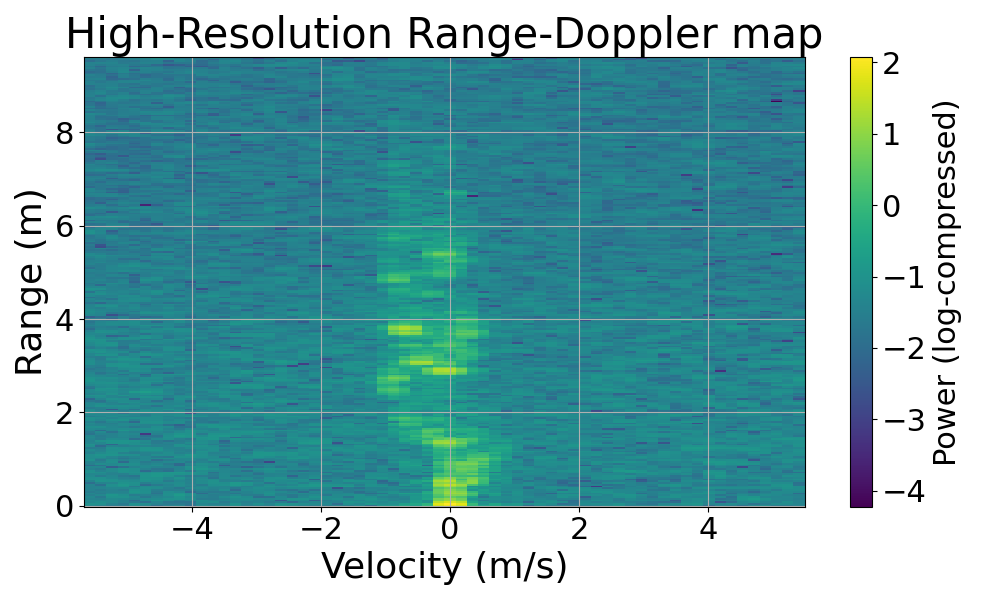}%
	}	
	\caption{Depiction of a) low-resolution rD map without zero-padding, b) low-resolution rD map with zero-padding, c) corresponding high-resolution rD map.}
	\label{fig:zeropadding}
\end{figure*}

Zero-padding before FFT was proposed for Doppler super-resolution in rD maps in \cite{rD_diffusion} and for azimuth super-resolution in range-azimuth maps in \cite{az_superres}. It is, however, not yet benchmarked against other methods to extend the rD map dimensions. We therefore compare the zero-padding version with the non zero-padding version, where $\mathbf{X}^{lr}_{t}\in\mathbb{R}^{(\frac{N_r}{2}+1)\times (\frac{N_D}{2})}$ and the extension of the rD map's range and Doppler dimensions is performed by pixel-shuffle \cite{pixel_shuffle} in the high-resolution rD map reconstruction block (Sec.~\ref{sec:extension}). An example of the same rD map computed with and without zero-padding before the range/Doppler FFTs is given in Fig.~\ref{fig:zeropadding}.

\subsection{Dimension shaping}
\label{sec:dim_shaping}
As we apply the range FFT to a real-valued baseband signal where the number of time steps is a power of two, the resulting number of non-negative frequency bins $N_r/2+1$ (without zero-padding) or $N_r+1$ (with zero-padding) is odd. We would like to use a number of bins that's also a power of two, as this simplifies upsampling. It also allows a straightforward future extension to a \gls{SWIN} U-Net \cite{swin_unet}, where the spatial signal dimension between blocks usually changes by factor two. We chose to drop the maximum range bin (Nyquist frequency), as we assume that it contains no relevant information and we don't need all non-negative (and complex-valued) frequency bins to convert back to time-domain via range \gls{iFFT}. We do the same with the high-resolution ground truth, such that $\mathbf{X}^{hr}_t$ has the shape $N_r\times N_D$. The output of dimension shaping is therefore $\mathbf{X}^{lr,d}_{t:t-3}\in\mathbb{R}^{\frac{N_r}{2}\times \frac{N_D}{2}\times 4}$ (without zero-padding) or $\mathbf{X}^{lr,d}_{t:t-3}\in\mathbb{R}^{N_r\times N_D\times 4}$ (with zero-padding). 

\subsection{Compression}
\label{sec:compression}
The received power of targets is proportional to the target's
\gls{RCS}, which varies greatly with target material, geometry, and viewing angle. In addition, the \gls{RCS} of targets usually fluctuates. The received power is further proportional to the angle-dependent antenna gains and to the inverse of the range to the power of four, given far-field conditions and two-way propagation \cite{richards_modern_radar}. Those received target returns may therefore have very different magnitudes (tens of decibels dynamic range). Directly using linear-scale inputs thus becomes a serious issue. The bottleneck is less underflow when during backpropagation using 32-bit or 64-bit floating point precision. The problem lies in weak signal components like rD bins of weak targets having negligible loss contribution when using losses like \gls{MSE} applied to linear-scale radar signals. Similarly, weak signal components are almost invisible at the network input in linear scale.

\begin{figure*}[h!]
	\centering
	\subfloat[Square-law: $\gamma=2$.]{%
		\includegraphics[width=0.3\textwidth]{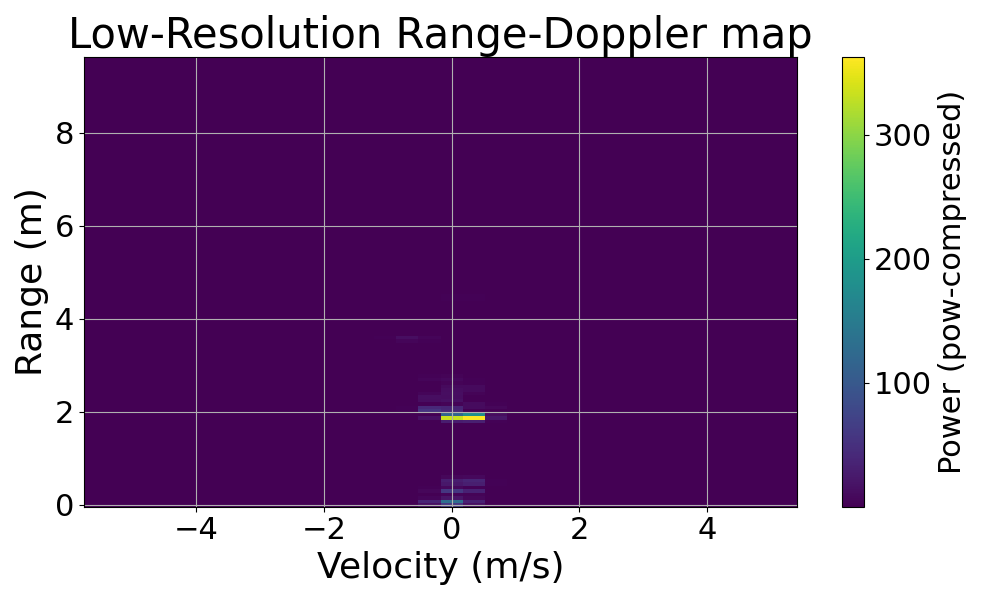}%
	}\hspace{0.03\textwidth}%
	\subfloat[Linear-scale: $\gamma=1$.]{%
		\includegraphics[width=0.3\textwidth]{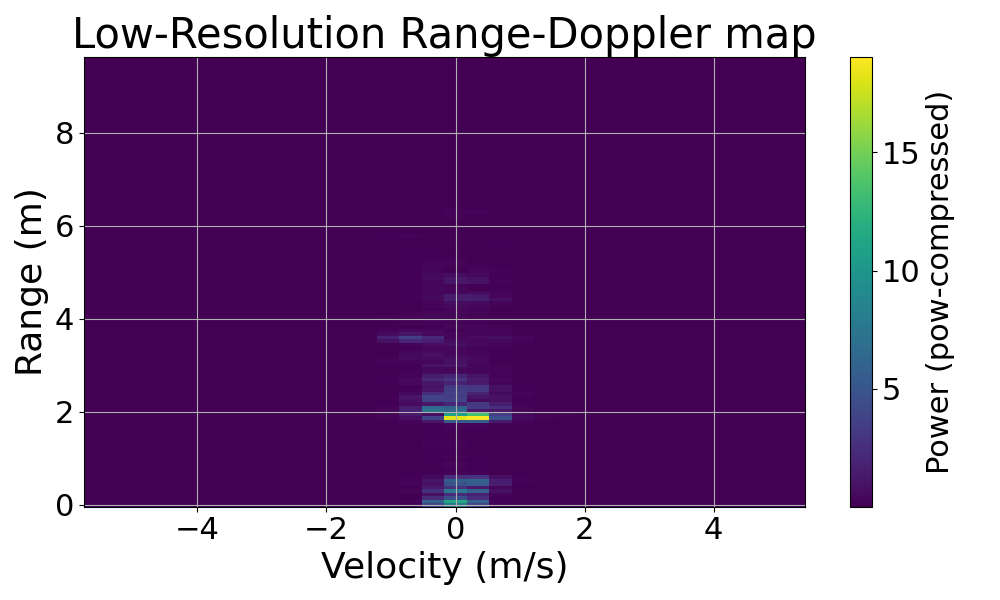}%
	}\hspace{0.03\textwidth}%
	\subfloat[Sqrt: $\gamma=0.5$.]{%
		\includegraphics[width=0.3\textwidth]{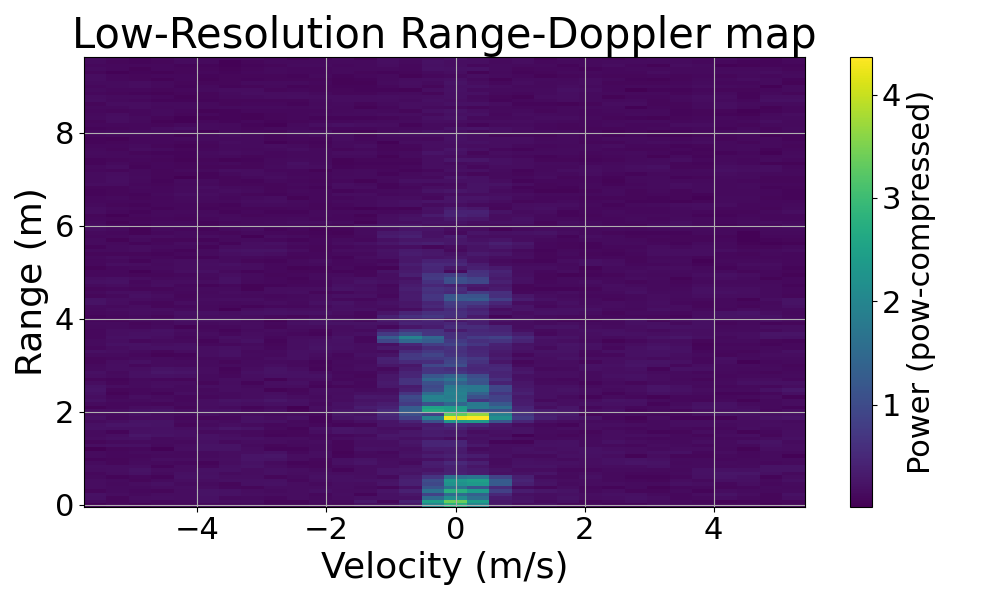}%
	}\\[0.5em]
	
	\subfloat[$\gamma=0.3$.]{%
		\includegraphics[width=0.3\textwidth]{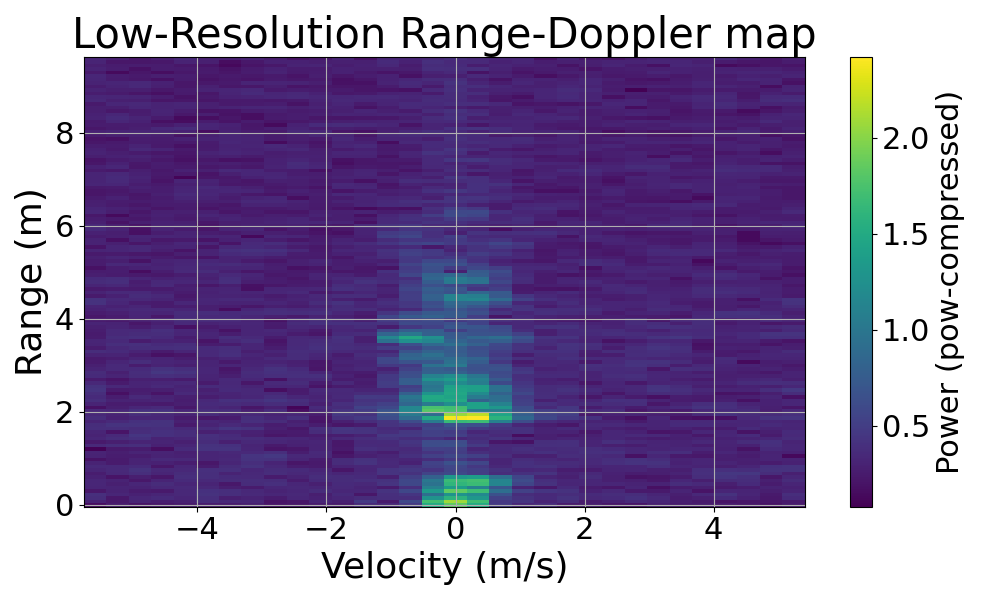}%
	}\hspace{0.03\textwidth}%
	\subfloat[$\gamma=0.1$.]{%
		\includegraphics[width=0.3\textwidth]{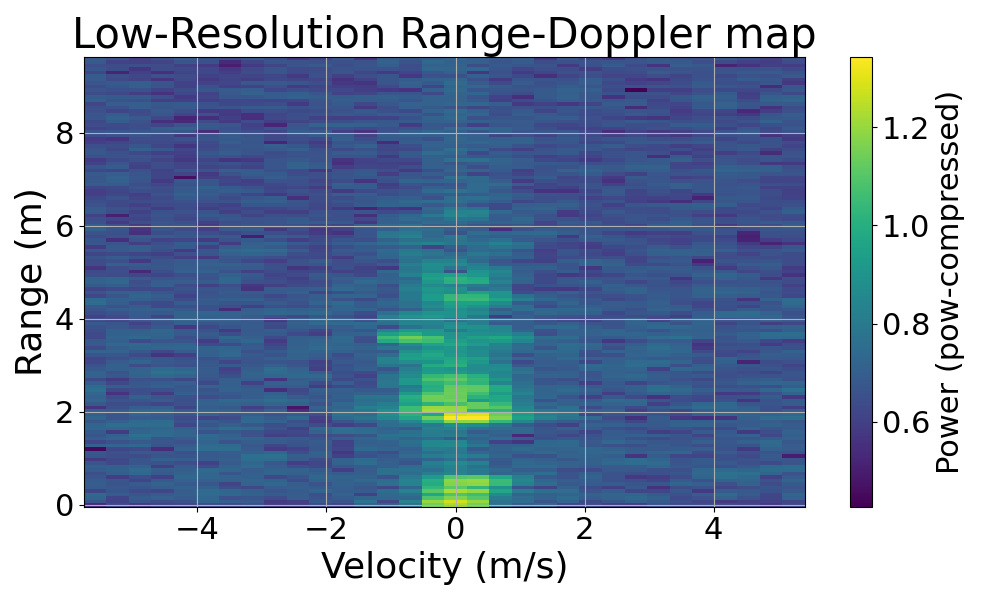}%
	}\hspace{0.03\textwidth}%
	\subfloat[$\log_{10}$]{%
		\includegraphics[width=0.3\textwidth]{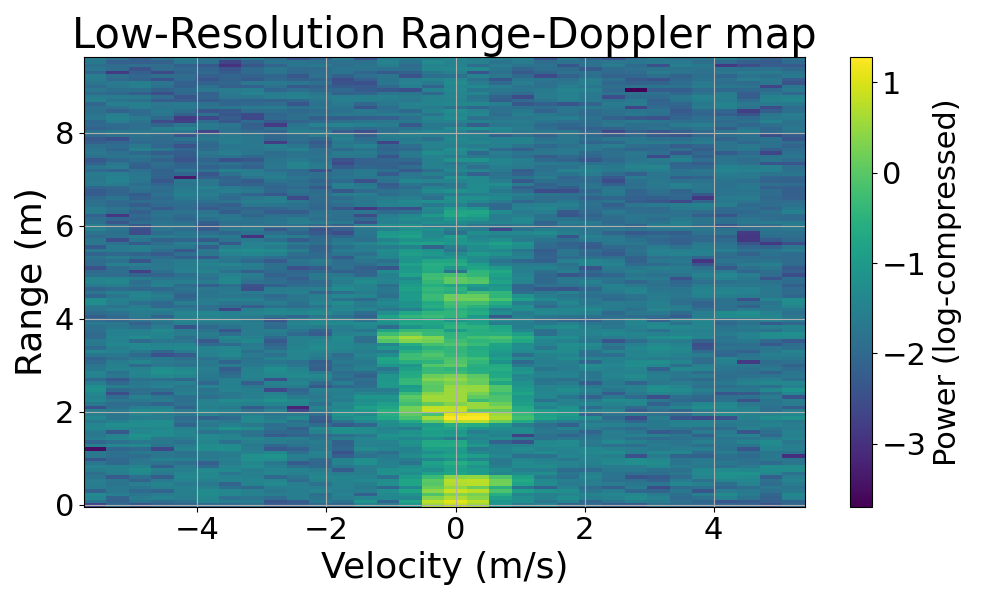}%
	}
	
	\caption{Comparison of different rD map representations (logarithmic or power-law compressed with factor $\gamma$) that can be used as network input and for loss computation.}
	\label{fig:compression}
\end{figure*}

To mitigate this problem, one can compress the rD bin magnitudes with a non-linear mapping, e.g. by applying a logarithm, a power-law compression, or a power scaling $\sim r^{4}$ as proposed in \cite{dp_radar_people_counting}. Non-linear compressions are best applied to non-negative magnitudes in frequency domain, as applying such functions to the real and imaginary parts of complex spectra distorts the signal's phase. Non-linear compression in time-domain distorts the spectrum and in case of complex I/Q signals also the phase. Non-linear frequency-domain magnitude compression is easily reversible. We test three magnitude compression variants $x_c = \comp(x_d)$, $x_d\in\mathbf{X}^{lr,d}$, $x_c\in\mathbf{X}^{lr,c}$, with
\begin{align}
	\comp_{lin}(x_d) &= x_d,\\
	\comp_{log}(x_d) &= \log_{10}\left(x_d\right),\\
	\comp_{pow}(x_d) &= (x_d)^\gamma,
\end{align}
i.e. no compression / linear (lin), log-compression with logarithm base 10 (log), and power-law compression (pow) with factor $\gamma$. Before compression, we lower-bound the spectrum magnitude with $x_d = \max(x_d, 10^{-12})$ to avoid numerical issues, as $\log_{10}(x_d)\rightarrow-\infty$ for $x_d\rightarrow0^+$. We give examples of the different compressions in Fig.~\ref{fig:compression}.

We believe, this representation aspect should be studied deeper in radar research. A hyperparameter search over different $\gamma$, stacking different representations, or learning the compression as done in learnable front-ends in audio \cite{trained_frontend_orig, ada_frontend_audio}, could likely bring significant performance gains. In this work, we investigate the simple variants (linear-scale, power-law, log-scale) to demonstrate the impact of a suitable magnitude compression.

For input normalization, we tried min-max normalization for each respective representation, where the min/max values were computed from the training set. However, with our chosen baseband signal amplification level, all representations except square-law already have suitable value ranges (see Fig.~\ref{fig:compression}). Adding min-max normalization worsened results, thus we give the compressed stack of rD map frames without additional input normalization to the next trainable network, the feature extractor.

\subsection{Feature extractor}
\label{sec:feat_extr}
The feature extractor is identical to the shallow feature extraction in SwinIR, i.e. a linear 2D convolutional layer \cite{dl_goodfellow} with $C$ kernels of size $(3\times 3)$ and stride one. It extracts a high-dimensional feature map $\mathbf{X}^{lr,e}\in\mathbb{R}^{N_r\times N_D\times C}$ from the compressed rD map frames $\mathbf{X}^{lr,c}_{t:t-3}\in\mathbb{R}^{N_r\times N_D\times 4}$ with
\begin{equation}
	\mathbf{X}^{lr,e} = \conv2d(\mathbf{X}^{lr,c}_{t:t-3})
\end{equation}
to serve as an improved input to the feature transformer.
\subsection{Feature transformer}
\label{sec:feat_transform}
\begin{figure}
	\centering
	\includegraphics[width=0.95\linewidth]{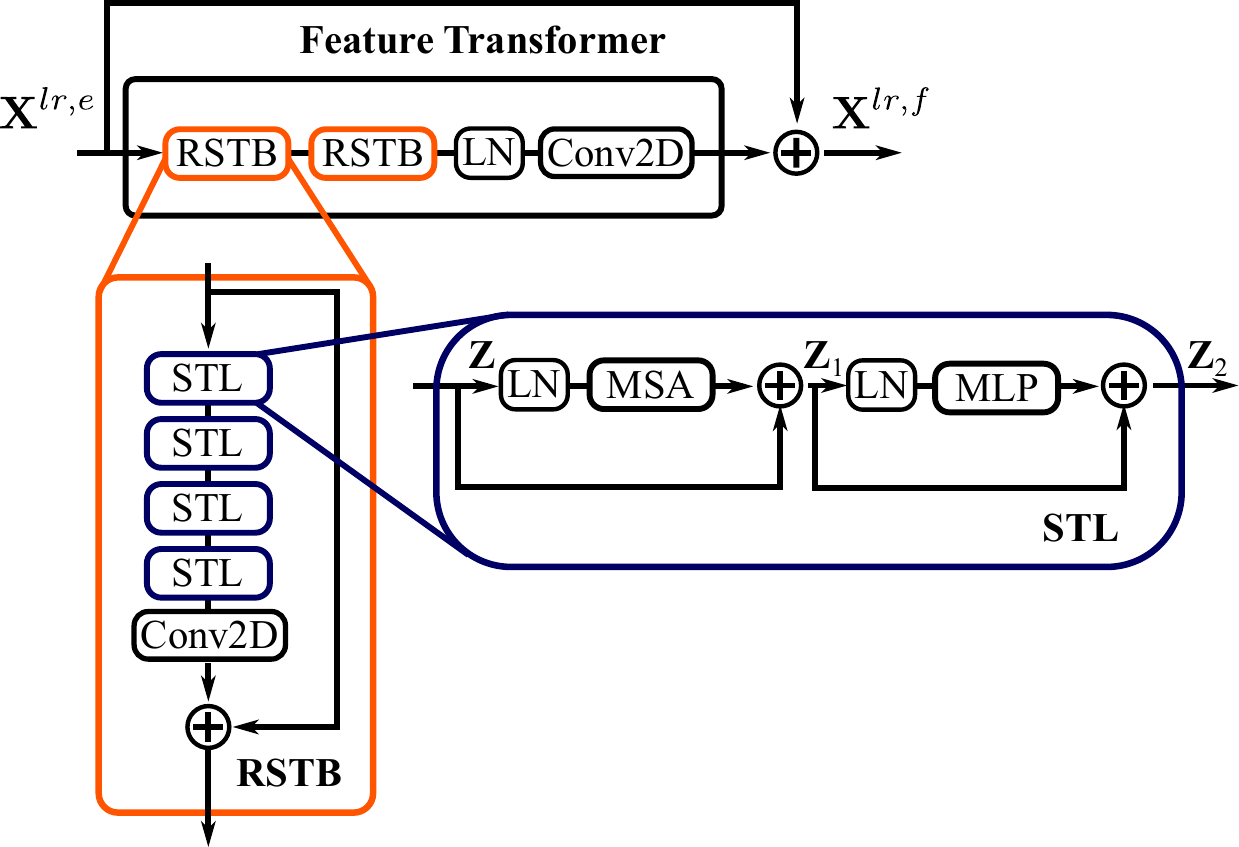}
	\caption{Depiction of the processing in the feature transformer. The channel-expanded input $\mathbf{X}^{lr,e}$ is processed by a series of \glspl{RSTB}, a 2D convolutional layer, and a residual addition, to output $\mathbf{X}^{lr,f}$. Each \gls{RSTB} contains a cascade of \glspl{STL}, a 2D convolutional layer, and a residual addition. The \glspl{STL} apply \gls{MSA} in parallel to non-overlapping windows of the split input. In consecutive \glspl{STL}, windows in the \gls{MSA} are shifted by half the window size to increase receptive field.}
	\label{fig:featuretransformer}
\end{figure}

The feature transformer is inspired by the deep feature extraction block in SwinIR, but differs in the number of layers (we used a smaller, fixed number of layers) and in the attention windows. The general structure is the same and depicted in Fig.~\ref{fig:featuretransformer}. It transforms the extracted rD map features into a better representation for the final rD super-resolution step. 

The feature transformer takes $\mathbf{X}^{lr,e}$ as input, which undergoes multiple \glspl{RSTB}, \gls{LN} \cite{layernorm}, and a final 2D convolutional layer with $C$ kernels of size $(3\times3)$ and stride one. The processed output  and input are then combined by residual addition, which gives $\mathbf{X}^{lr,f}\in\mathbb{R}^{N_r\times N_D\times C}$ with
\begin{equation}
	\mathbf{X}^{lr,f} = \conv2d(\LN(\mathrm{RSTB}_m(\mathbf{X}^{lr,e}))) + \mathbf{X}^{lr,e},
\end{equation}
where $\rstb_m$ is a cascade of $m$ \glspl{RSTB}. In this work, we set $m=2$.
Each \gls{RSTB} contains a cascade of $n=4$ \glspl{STL}, followed by another 2D convolutional layer with kernel size $(3\times3)$ and stride one, and residual addition. Finally
\begin{equation}
\rstb_{m}(\mathbf{Z}) = \conv2d(\stl_n(\mathbf{Z})) + \mathbf{Z},
\end{equation}
where $\stl_n$ is a cascade of $n$ \gls{STL} blocks.
In each \gls{STL}, the input $\mathbf{Z}$ first goes through \gls{LN}. It is then split into non-overlapping windows. A weight-shared SWIN transformer encoder, which applies \gls{MSA} with learnable relative positional encoding \cite{swin}, then processes all windows in parallel and the processed windows are stacked back into the shape of the input $\mathbf{Z}$. This is followed by residual addition, \gls{LN}, a \gls{MLP}, and another residual addition
\begin{equation}
	\mathbf{Z}_1 = \mathrm{MSA}(\LN(\mathbf{Z}))+\mathbf{Z},
\end{equation}
\begin{equation}
	\mathbf{Z}_2 = \mathrm{MLP}(\LN(\mathbf{Z}_1))+\mathbf{Z}_1.
\end{equation}
The weight-shared \gls{FFW} network is applied point-wise to each rD bin. It has a \gls{GELU} activation \cite{gelu} after the first layer. In our implementation, the first layer has $2C$ neurons for channel expansion, followed by the second layer with $C$ neurons.
SwinIR uses SWIN attention with 2D windows. Our implementation can also use other windows as described below.

\textbf{SWIN attention:}
\gls{SWIN} transformer with \gls{SWIN} attention \cite{swin} has been introduced as a leight-weight alternative to \glspl{ViT} \cite{vit}. While \glspl{ViT} are found in very large models as vision backbone \cite{clip} due to their global attention mechanism, \gls{SWIN} is popular with less data or hardware restrictions. 

\begin{figure}
	\centering
	\includegraphics[width=0.7\linewidth]{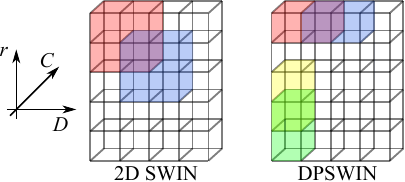}
	\caption{Depiction of the windowing in the \gls{STL} blocks of an \gls{RSTB}. For 2D SWIN (left), 2D rD windows are used. In each \gls{STL} block, a window shift happens, where windows shift back and forth. In \gls{DPSWIN}, 1D windows are applied separately along both axes, first in range (yellow/green windows), then in Doppler (red/blue windows). For \gls{DP}, 1D windows that span over a full axis are applied separately along range and Doppler.}
	\label{fig:swins}
\end{figure}

In \gls{SWIN} attention applied to \gls{rD} maps, the rD map is first split into equally sized, non-overlapping, square 2D windows of size $(W\times W)$. A flattened $(W\times W)$ window gives $W^2$ tokens of dimension $C$. A transformer encoder with learnable relative positional encoding then applies attention independently within each window. To share information between windows and to increase the receptive field of neurons in deeper layers, windows are shifted back and forth between \glspl{STL} by half the window size. The process for 2D SWIN is depicted in Fig.~\ref{fig:swins} (left).

2D windows are well suited for images with rich spatial structure in both dimensions. 1D shifted windows were later proposed for time domain \gls{ECG} processing \cite{swin_1D_time_series}, for target recognition in high-resolution range-profiles \cite{radar_swin_hrr}, and for super-resolution of high-resolution range profiles, including a complex-valued version \cite{swin_1D_hrr}. There's also rectangular windows applied to super-resolution of \gls{LiDAR} point clouds \cite{lidar_upsampling_rect_swin}.

\textbf{Axial \& dual-path attention:}
\label{sec:dp}
\gls{DP} attention \cite{dptnet, SepFormer} is a special case of axial attention \cite{axial_attention} for 2D signals. It is popular in audio and has later also been used in radar \cite{irs22_radar_separation, trans_radar}. As \gls{SWIN}, it applies local attention, but to 1D axial windows that span over a whole axis without any window shifting, i.e. 1D windows over the range or Doppler axis, respectively. By applying axial attention along both axes (dual-path) and combining the outputs, global attention over the whole input is achieved.

With axial processing applied to \gls{rD} maps, we leverage the knowledge that 1D \glspl{FFT} are applied independently along the range and Doppler axes. For a point target, this produces a 2D sinc-shaped target spectrum that is separable as a product of two 1D sinc functions. Efficient 1D processing might therefore be sufficient for rD super-resolution. The axial rD map processing is depicted in Fig.~\ref{fig:swins} (right) for the more general case of \gls{DPSWIN} described next.

\textbf{Dual-path SWIN (DPSWIN) attention:}
\label{sec:swin}
In this work, we fuse the idea of axial/dual-path attention with 1D \gls{SWIN} attention and apply 1D shifted windows separately along the range and Doppler axes to reduce computational cost, as shown in Fig.~\ref{fig:swins} (right). 

In the 2D square windows of \gls{SWIN}, the computational complexity of the attention matrix multiplications is \cite{swin}
\begin{equation}
	\Omega(\mathrm{SWIN}) = 2W^2N_rN_DC,
\end{equation}
which is proportional to the squared window size $W^2$ instead of $(N_rN_D)^2$ for global attention. It is the product of the cost of both matrix multiplications in the attention per window $2(W^2)^2C$ and the number of windows $(N_rN_D)/W^2$.
In a dual-path transformer, the cost of the attention matrix multiplications per window is $2N_r^2C$ for range windows and $2N_D^2C$ for Doppler windows with a number of $N_D$ windows in range and $N_r$ windows in Doppler. The average computational cost per axis is
\begin{equation}
	\Omega(\mathrm{DP})=(N_r+N_D)N_rN_DC,
\end{equation}
which is problematic for high $N_r$ or $N_D$.

In \gls{DPSWIN}, the attention matrix multiplication cost per 1D window is $2W^2C$ and the number of 1D windows is $(N_rN_D)/W$. The computational cost of the attention matrix multiplications in \gls{DPSWIN} attention is  
\begin{equation}
	\Omega(\mathrm{DPSWIN}) = 2WN_rN_DC,
\end{equation}
i.e. a reduction to linear complexity in window size because of 1D not 2D windows. Radar spectra are often sparse, where local windows are sufficient, large objects can be captured by choosing cheaper 1D windows of increased size, and receptive field can be increased by stacking such \gls{DPSWIN} layers. This design therefore scales very well to high-dimensional input. For our case it might also help in resolving closely spaced objects, as large 1D windows can still capture the full sinc-shaped target pattern, whereas 2D analogues would be too expensive. We evaluate \gls{DPSWIN} w.r.t. 2D \gls{SWIN} attention and implicitly w.r.t. dual-path attention, which is a special case of \gls{DPSWIN} attention where the window goes over the full axis and thus no window shifting occurs. One disadvantage of such \gls{DPSWIN} attention is that four instead of two layers are required to increase receptive field in both dimensions, i.e. two STLs for overlapping shifted windows to propagate information in range and equivalently two STLs for Doppler. This motivates our choice of four serial \glspl{STL} in a \gls{RSTB} block, two per axis, as shown in Fig.~\ref{fig:featuretransformer}.
\subsection{High-resolution range-Doppler map reconstruction}
\label{sec:extension}
If the signal was not zero-padded in the rD processing, its range and Doppler dimensions are extended in this module with pixel-shuffle \cite{pixel_shuffle}. Pixel-shuffle reshapes a $(N_r/2\times N_D/2\times C)$ low-resolution rD feature map into a $(N_r\times N_D\times C/4)$ super-resolution rD feature map. We do the spatial extension in feature space and only compress the channel dimension to one rD map output channel in a subsequent convolutional layer to facilitate training by de-coupling spatial extension and rD map estimation. A final convolutional layer may also compensate for possible artifacts created by the spatial channel reordering. The compressed super-resolution rD map $\mathbf{X}^{sr}_t\in\mathbb{R}^{N_r\times N_D}$ corresponding to $\mathbf{Y}^{lr}_t$ is 
\begin{equation}
	\mathbf{X}^{sr}_t = \conv2d(\pixelshuffle(\mathbf{X}^{lr,f})),
\end{equation}
where the 2D convolutional layer has one kernel of size $(3\times3)$. With zero-padding, we skip the pixel-shuffle and apply the last convolutional layer to a $\mathbf{X}^{lr,f}\in\mathbb{R}^{N_r\times N_D\times C}$ feature map. As the channel dimension is $C$ not $C/4$ before the final convolutional layer, a model using zero-padding has a few negligible extra parameters. Finally, an optional softplus activation \cite{dl_goodfellow} is added to enforce network outputs $\geq 0$ when using linear or power-law compressed magnitudes. Alternatively, we round them to zero in post-processing.

We drop the time index $t$ and define a \gls{RMSE} loss $\mathcal{L}_{RMSE}$ in this compressed rD map as the \gls{RMSE} between the compressed high-resolution rD map $\mathbf{X}^{hr}$ and the compressed super-resolution rD map $\mathbf{X}^{sr}$
\begin{equation}
	\mathcal{L}_{RMSE}=\sqrt{\langle\left(\mathbf{X}^{hr}-\mathbf{X}^{sr}\right)^2\rangle},
	\label{eq:trainrmse}
\end{equation}
where $\langle \,\cdot\, \rangle$ denotes the arithmetic mean of all rD bins. We chose \gls{RMSE}, as it matches the \gls{LSD} definition \eqref{eq:lsd_loss}, which is defined by log-compressed rD bin magnitudes that we employ as effective compression baseline. We kept the square root for the other compressions for consistency. Opposed to the more standard \gls{MSE} loss, the additional square root compression after aggregation increases the impact of weaker signals in training, which might improve results with weak compressions where the dynamic range of compressed magnitudes is larger.

\section{CFAR detection and CFAR losses}
\label{sec:cfar}
In this section, we shortly review \gls{CFAR} detection for rD maps. For a deeper handling of radar target detection theory, we refer to \cite{richards_modern_radar}. By extending the work in \cite{cfar_loss}, we then show how one can derive differentiable \gls{CFAR} losses that are based on soft \gls{CFAR} threshold differences between high-resolution rD maps $\mathbf{X}^{hr}_{t}$ and super-resolution \gls{rD} maps $\mathbf{X}^{sr}_{t}$. We argue that adding such \gls{CFAR} loss aligns the downstream target detection behavior. Thus, we introduce various \gls{CA-CFAR} loss variants and study their impact when added to the \gls{RMSE} loss.
\subsection{CFAR detection}
In radar, \gls{CFAR} detectors are used for target detection in the presence of interference with optimized probability of detection $P_D$, constrained by a given probability of false-alarm $P_{FA}$.

\gls{CFAR} detectors extend Neyman-Pearson detection to deal with non-stationary clutter. \gls{CFAR} detectors therefore adaptively change the detection threshold that discriminates \gls{rD} bins containing only interference (null hypothesis $\mathcal{H}_0$) and rD bins containing target-plus-interference ($\mathcal{H}_1$ hypothesis). Most often, a square-law detector is used, i.e. the squared magnitude spectrum $|\mathbf{X}|^2$ is the \gls{CFAR} input. Applied to 2D rD maps, the \gls{CFAR} detector compares the power in each complex \gls{rD} bin $x_{k,l}\in\mathbb{C}$ with a threshold $T$. The respective tested rD bin is called the \gls{CUT} and the discrimination into $\mathcal{H}_0$ and $\mathcal{H}_1$ is given as

\begin{equation}
	|x_{k,l}|^2 \underset{\mathcal{H}_0}{\overset{\mathcal{H}_1}{\gtrless}} T,\,\,\textrm{with}\,\,\, T = \alpha\hat{\sigma}^2.
\label{eq:tresholding}
\end{equation}
The parameter $\alpha$ is the \gls{CFAR} constant that ensures the defined $P_{FA}$ and $\hat{\sigma}$ is the interference statistic. In our work, we only study \gls{CA-CFAR}, where 
\begin{equation}
	\alpha = N\left(P_{FA}^{-1/N}-1\right),
\end{equation}
with $N=|\mathcal{N}|$ being the number of rD bins $\mathcal{N}$ used in the computation of the interference statistic (the so called reference cells). The \gls{CA-CFAR} interference statistic is given by
\begin{equation}
	\hat{\sigma}^2 = \frac{1}{N} \sum_{(k,l)\in\mathcal{N}}|x_{k,l}|^2.
\end{equation}
The reference cells $\mathcal{N}$ contain the rD bins in a 2D band around the respective \gls{CUT}. Since target power might be distributed among multiple neighboring rD bins, thus biasing the interference statistic, usually the closest rD bins around the \gls{CUT}, the guard cells, are not part of the reference cells for interference statistic computation. The size of the guard cell band around the \gls{CUT} and of the reference cell band around the guard cells, as well as $P_{FA}$, are often tuned by hand.\footnote{The shapes of the sets of guard and reference cells can be arbitrary and don't have to be square-shaped bands.} To avoid incomplete bands, we apply the standard method of mirroring rD bins at the border of the rD map.

The interference statistic computation is an arithmetic mean for \gls{CA-CFAR}, thus differentiable, and $\alpha$ is a constant scaling factor. The only non-differentiable operation is the hard thresholding in \eqref{eq:tresholding}. We next describe how one can derive differentiable \gls{CFAR} losses based on this knowledge, as previously done for one of the following \gls{CFAR} losses in \cite{cfar_loss}. For other \gls{CFAR} variants with non-differentiable interference statistic computation, one might replace only the non-differentiable operator (e.g. ranking in \gls{OS-CFAR}) with a known differentiable approximation, or approximate the interference statistic computation with a pre-trained or adversarially learned neural network.

It should be noted that the underlying assumptions for \gls{CA-CFAR} such as reference cells being free from targets, or containing the same i.i.d. complex Gaussian noise as the \gls{CUT} \cite{richards_modern_radar}, are violated in our scenarios. The radar frontend and Tx/Rx coupling might introduce non-Gaussian noise. Reflected power by different persons may lie in the same rD map neighborhood, and strong clutter by the ground, walls, street signs etc., is correlated and non-uniform.
For readers interested in how one might use deep learning to achieve \gls{CFAR} behavior in complex environments, we refer to \cite{cfarnet}. However, our motivation is to enforce the same \gls{CFAR} detector behavior with our super-resolution network output $\mathbf{X}^{sr}_t$ as with the real high-resolution rD map $\mathbf{X}^{hr}_t$. Even if the detector does not have the property of \gls{CFAR} in practice, penalizing wrong CFAR detector behavior might still lead to higher quality super-resolution rD maps by enforcing correct local target-plus-interference and interference statistics.  

\subsection{CA-CFAR losses with differentiable interference statistic}
\begin{figure*}[h]
	\centering
	\subfloat[Log-compressed rD map.]{%
		\includegraphics[width=0.3\textwidth]{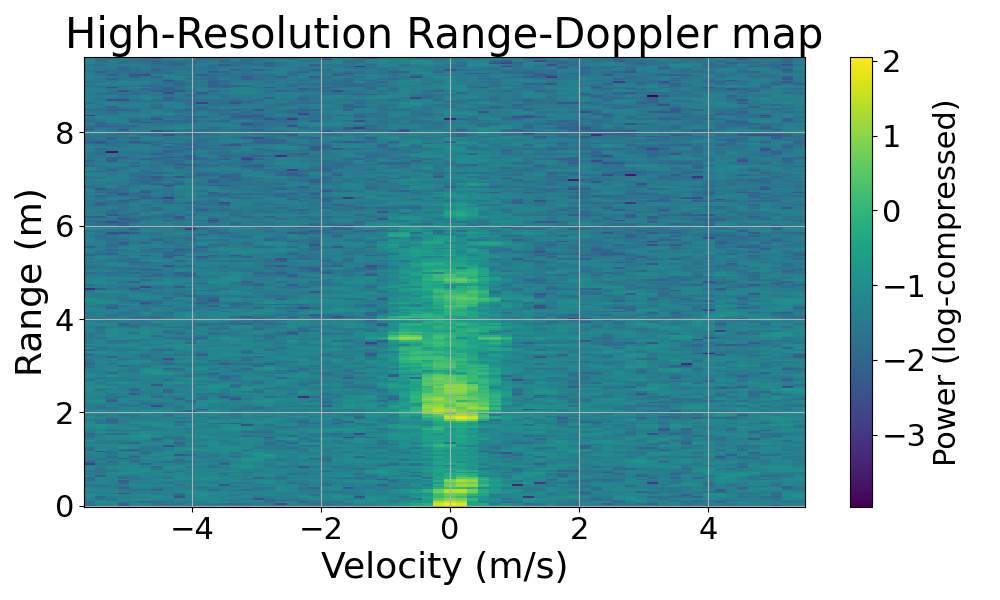}%
	}\hspace{0.03\textwidth}%
	\subfloat[Soft mask with $t^1$.]{%
		\includegraphics[width=0.3\textwidth]{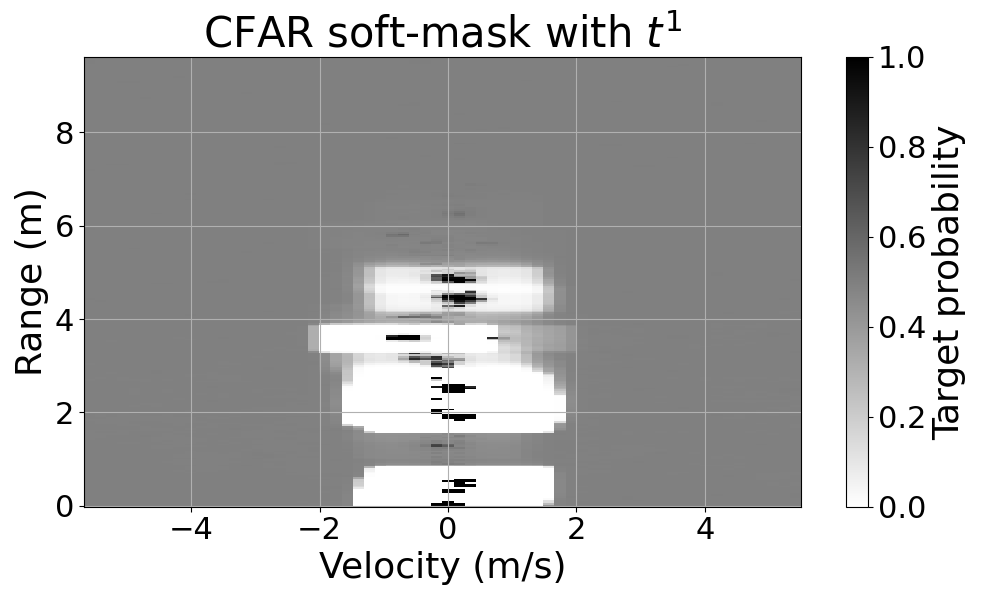}%
	}\hspace{0.03\textwidth}%
	\subfloat[Soft mask with $t^2$.]{%
		\includegraphics[width=0.3\textwidth]{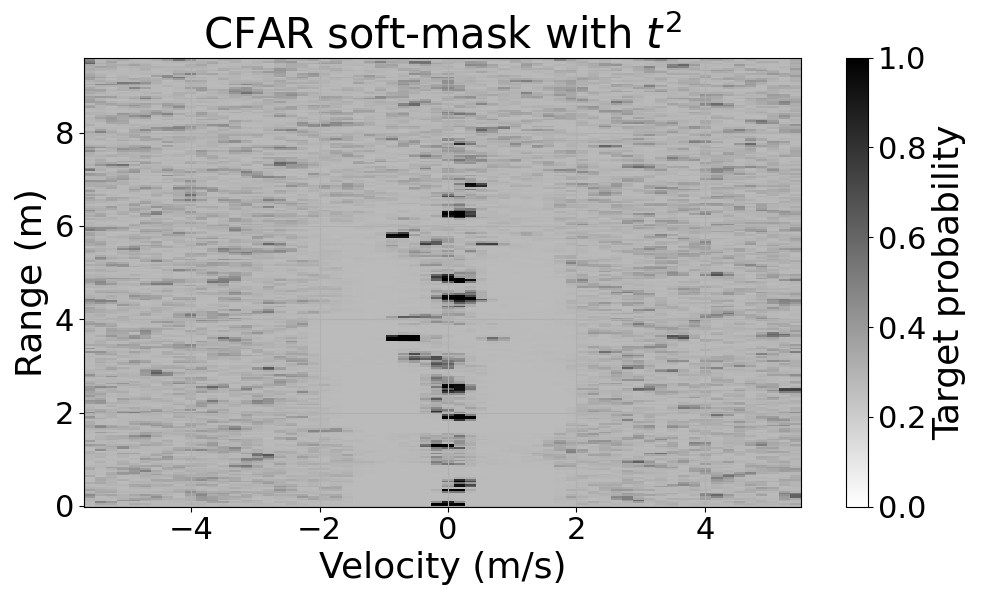}%
	}\\[0.5em]
	
	\subfloat[Soft mask with $t^3$.]{%
		\includegraphics[width=0.3\textwidth]{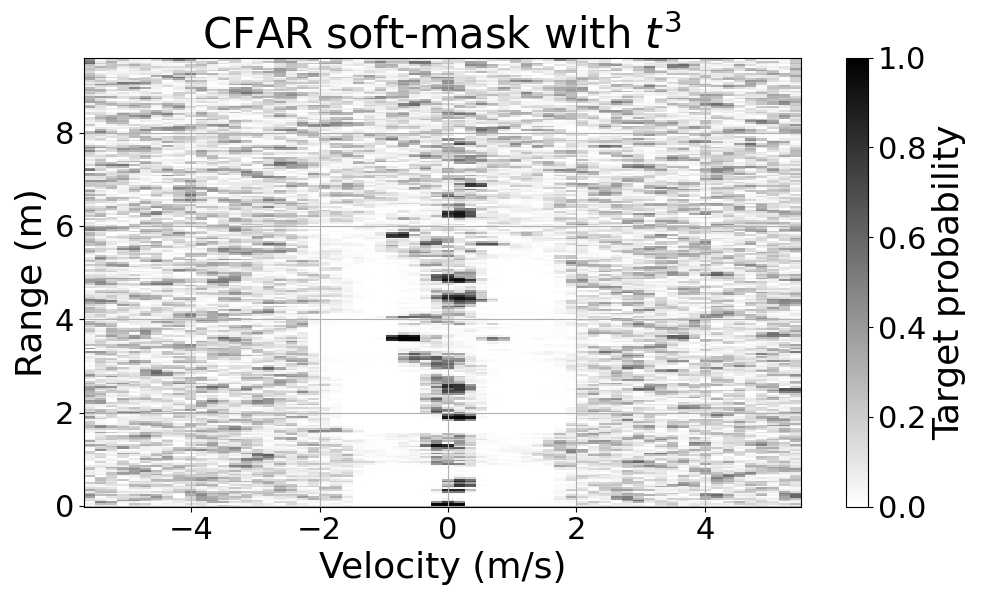}%
	}\hspace{0.03\textwidth}%
	\subfloat[Soft mask with $t^3$ and $\xi=10$.]{%
		\includegraphics[width=0.3\textwidth]{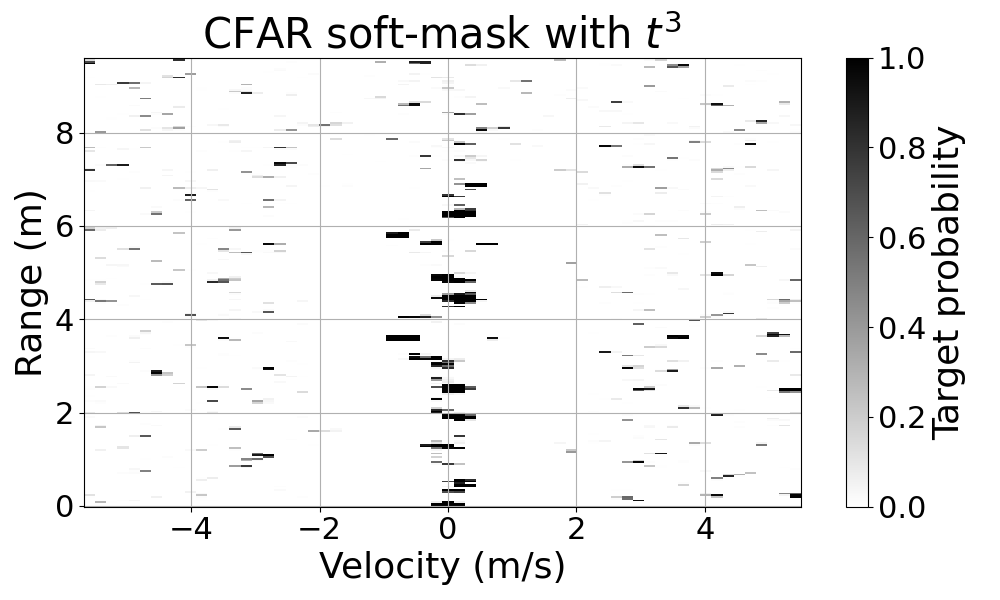}%
	}\hspace{0.03\textwidth}%
	\subfloat[Hard mask.]{%
		\includegraphics[width=0.3\textwidth]{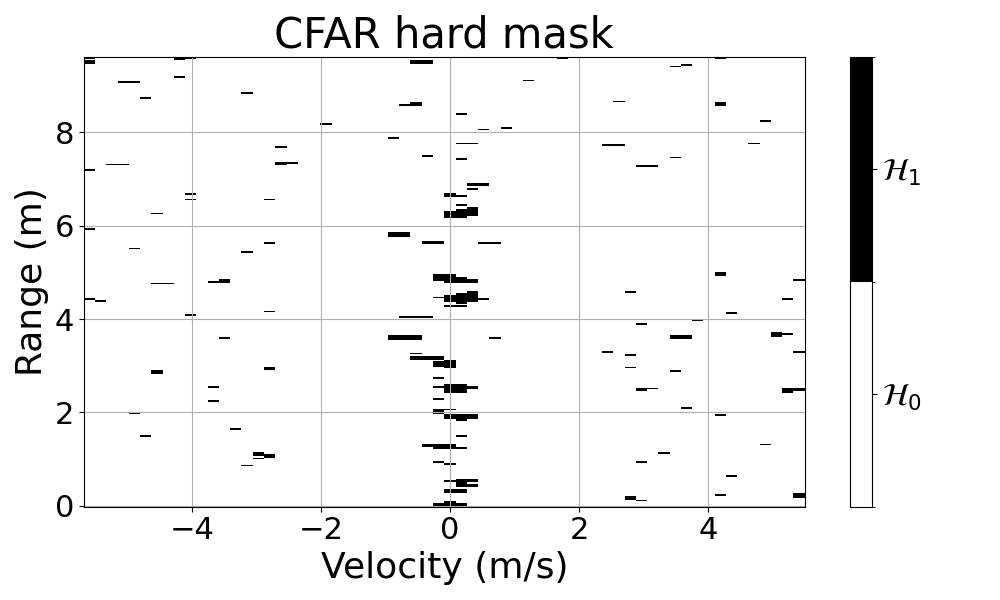}%
	}	
	\caption{Comparison of different CFAR masks for the rD map in a) with $\xi=1$ in b)-d) and with $\xi=10$ in e). All masks are computed with $P_{FA}=10^{-2}$, a square-shaped guard cell band of width two cells around the \gls{CUT}, and a square-shaped reference cell band of width five cells around the guard cells.}
	\label{fig:cfar_masks}
\end{figure*}

A non-differentiable hard threshold can be replaced by a soft threshold with a differentiable sigmoid  function\cite{dl_goodfellow}, where an additional hyperparameter $\xi$ defines the steepness of the sigmoid. We denote the soft-threshold output by the sigmoid as a mask value $m_{k,l}$ with
\begin{equation}
	m_{k,l} = \sigmoid(\xi t_{k,l}).
\end{equation}
The second term $t_{k,l}$ in the sigmoid argument can be defined in different ways. By subtracting $T$ from \eqref{eq:tresholding}, the first proposal for $t_{k,l}$ becomes
\begin{equation}
	t^1_{k,l} = |x_{k,l}|^2-T.
\end{equation}
In this version, the difference between the \gls{CUT} power and the threshold is computed. This drives the sigmoid into saturation for large power difference and is likely suboptimal. Further dividing by $T$ gives 
\begin{equation}
	t^2_{k,l} = \frac{|x_{k,l}|^2}{T}-1.
\end{equation}
The authors of \cite{cfar_loss} proposed and used an expression equivalent to $t^2_{k,l}$. The issue with $t^2_{k,l}$ is that the sigmoid saturates at $\sigmoid(-\xi)$ for zero \gls{CUT} power, thus it cannot produce low mask values if $\xi$ is not set very high. A steep sigmoid with high $\xi$ is sensitive to small perturbations around zero, i.e. if $|x_{k,l}^2|\approx T$. The associated high gradients can degrade training stability. A steep sigmoid also saturates earlier, which may slow down convergence.
Adding 1 and applying a natural logarithm leaves
\begin{equation}
	t^3_{k,l} = \log\!\left(\frac{|x_{k,l}|^2}{T}\right).
\end{equation}
This log-ratio makes the mask value symmetric w.r.t. the $\frac{|x_{k,l}|^2}{T}$ power ratio, i.e. if $\frac{|x_{k,l}|^2}{T}$ gives $m_{k,l}$, then its reciprocal gives $1-m_{k,l}$. $t^3$ is therefore not only numerically more stable, but also physically and probabilistically more grounded. We compare $t^1$, $t^2$ and $t^3$ in Fig.~\ref{fig:cfar_masks}.

A \gls{CFAR} loss can then be defined based on the deviation between the \gls{CFAR} mask $m_{k,l}$ of a high-resolution rD map and a corresponding \gls{CFAR} mask $\hat{m}_{k,l}$ of a super-resolution rD map. By interpreting each \gls{CFAR} mask bin as an independent Bernoulli random variable denoting the target probability (see Fig.~\ref{fig:cfar_masks}), the problem becomes bin-wise binary classification and we apply the average bin-wise \gls{BCE} loss \cite{cfar_loss}
\begin{equation}
	\begin{split}
		\mathcal{L}_{CFAR}^{BCE} ={}&
		-\frac{1}{N_rN_D}\sum_{k=1}^{N_r}\sum_{l=1}^{N_D}
		m_{k,l}\log(\hat{m}_{k,l}) \\
		&-\frac{1}{N_rN_D}\sum_{k=1}^{N_r}\sum_{l=1}^{N_D}
		(1-m_{k,l})\log(1-\hat{m}_{k,l}).
	\end{split}
\end{equation}
We have further treated the mask as a categorical distribution over the rD bins and minimized the \gls{KLD} between both normalized mask distributions with $m'$ and $\hat{m}'$ such that $\sum m'_{k,l} = \sum \hat{m}'_{k,l} = 1$. This gives
\begin{equation}
	\mathcal{L}_{CFAR}^{KLD} = \frac{1}{N_rN_D}\sum_{k=1}^{N_r}\sum_{l=1}^{N_D} m'_{k,l}\log\left(\frac{m'_{k,l}}{\hat{m}'_{k,l}}\right).
\end{equation}
As last option, we treated it as regression and minimized \gls{MSE}
\begin{equation}
	\mathcal{L}_{CFAR}^{MSE} = \frac{1}{N_rN_D}\sum_{k=1}^{N_r}\sum_{l=1}^{N_D} (m_{k,l} - \hat{m}_{k,l})^2.
\end{equation}
With $t^2$ and $t^3$, the \gls{CFAR} losses are invariant to arbitrary equal scaling of target and interference power. Invariance to equal scaling of signal and interference power is a requirement for the CFAR property to hold. $t^1$ violates this power invariance property. 

\section{Experiments}
\label{sec:exp}
\subsection{Dataset}
Due to the limited number of works on deep learning-based rD map super-resolution, especially for both dimensions, currently no benchmark datasets exist for this task. This makes comparisons with the \gls{SOTA} infeasible. We therefore recorded our own dataset, which will be made openly available upon publication.

It is collected with our hand-held radar in pedestrian settings in Stuttgart, Germany. Roughly half of the dataset is recorded in outdoor environments including different public squares, (shopping) streets, and a wine festival. The other half is collected in indoor environments or partly closed spaces including the university library, lecture buildings, a pedestrian tunnel, and different subway stations and underground areas below the main station. We recorded with varying walking speeds. A few measurements were also taken at slow speed on a bicycle. We collected different levels of pedestrian traffic ranging from no persons in the recordings to clouds of people, e.g. in crowded subway stations or in the wine festival.

The sizes of the training, validation, and test sets are given in Tab.~\ref{tab:dataset}. The train/val/test split was done such that they contain completely independent environments or large differences, e.g. different zones of the same shopping street.

\begin{figure*}[h!]
	\centering
	\subfloat[Pedestrians in a shopping street.]{%
		\includegraphics[width=0.3\textwidth]{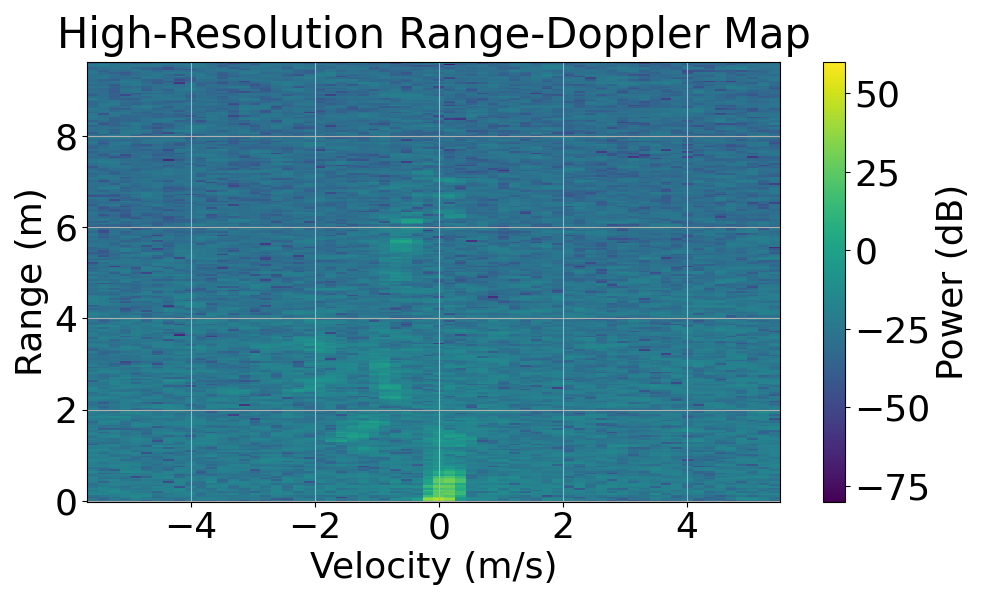}%
	}\hspace{0.03\textwidth}%
	\subfloat[Clouds of people between tents in a wine festival.]{%
		\includegraphics[width=0.3\textwidth]{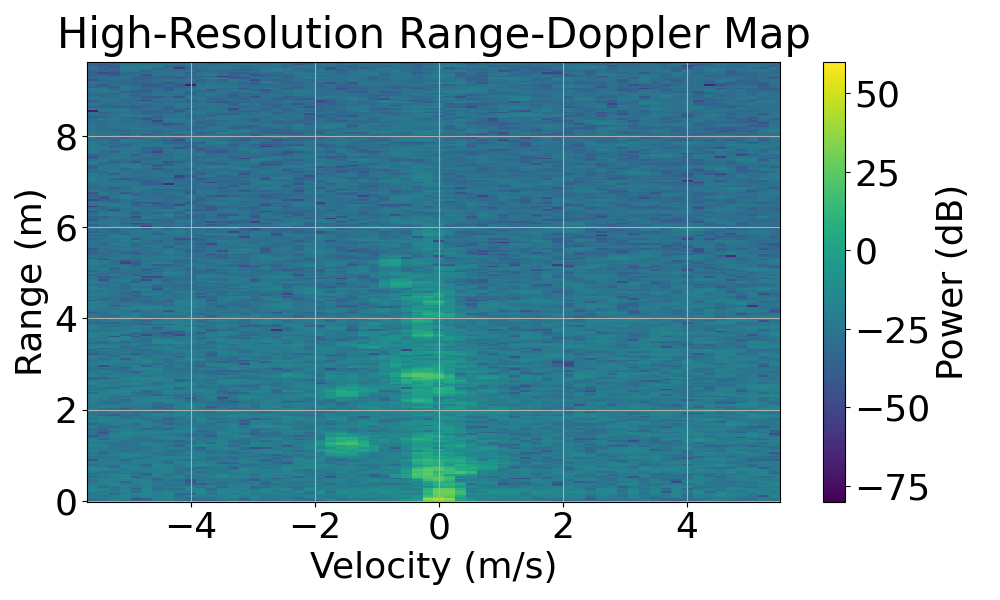}%
	}\hspace{0.03\textwidth}%
	\subfloat[A subway station with pedestrians, walls, pillars etc.]{%
		\includegraphics[width=0.3\textwidth]
		{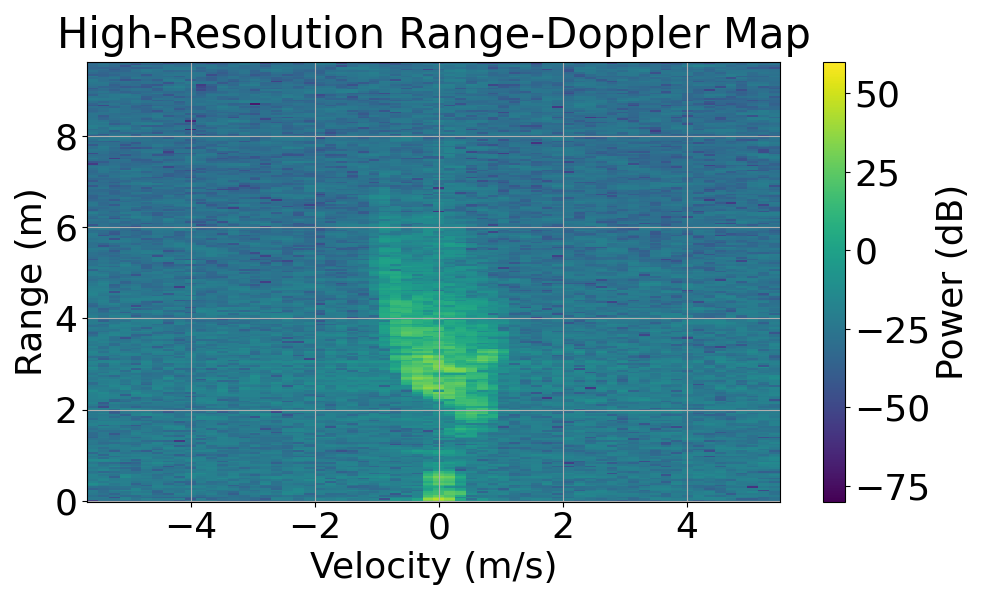}%
		\label{fig:dataset3}
	}	
	\caption{Different scenes in the dataset.}
	\label{fig:dataset}
\end{figure*}

Each sample holds a stack of 4 consecutive high-resolution 2D time-domain frames. Each frame is separated by $\SI{100}{ms}$. We record 8 consecutive frames to create two 4-frame samples. Between each 8-frame stack in a recording, we take a break of around \SI{1.2}{s} to avoid redundant, highly correlated samples.
\begin{table}[t]
	\centering
	\caption{Dataset sizes}
	\label{tab:dataset}
	\begin{tabular}{l c c c}
		\toprule
		Dataset & Train & Val  & Test  \\
		\midrule
		Indoor samples    & 3184 & 740 & 870 \\
		Outdoor samples  & 3034 & 706 & 752 \\
		\bottomrule
	\end{tabular}
\end{table}
The test set was only used for final evaluation. We give examples of recorded rD maps to show the variance in the recorded data. For following visualizations including the dataset samples in Fig.~\ref{fig:dataset}, we always use log-spectrum $10\log_{10}\left(|\mathbf{X}|^2\right)$. Due to privacy laws, we only recorded rD maps without corresponding images.

\subsection{Radar parametrization}
We set the parameters of the chirp-sequence such that a good trade-off in maximum range $r_{max}$, maximum velocity $v_{max}$, range resolution $\Delta r$, and velocity resolution $\Delta v$ for the pedestrian setting is reached. The key parameters are given in Tab.~\ref{tab:chirp_sequence}. Note that due to the asymmetry in $N_r$ and $N_D$, attention windows that go over the full Doppler axis can still be shifted in range.
\begin{table}[t]
	\centering
	\caption{High-resolution chirp-sequence radar parameters and corresponding range and Doppler resolutions.}
	\label{tab:chirp_sequence}
	\begin{tabular}{c c c c c}
		\toprule
		$f_c$ & $B$ & PRT & $N_r$ & $N_D$ \\
		\midrule
		61$\,$GHz & 4$\,$GHz & 220$\,\mu$s & 256 & 64 \\
		\bottomrule
	\end{tabular}

	\vspace{.4em}
	\begin{tabular}{c c c c}
		\toprule
		$v_{max}$ & $r_{max}$ & $\Delta v$ & $\Delta r$ \\
		\midrule
		5.59\,m/s & 9.6\,m & 0.175\,m/s & 0.0375\,m \\
		\bottomrule
	\end{tabular}
\end{table}

\subsection{Network architecture, training, and FFT window}
Our processing is implemented in TensorFlow 2 \cite{tensorflow2015-whitepaper}. All \glspl{FFT} use Hamming windows \cite{fft_windows} spanning over all non zero-padded fast-time or slow-time samples of the respective low-resolution and high-resolution time-domain frames. 

We performed initial investigations to find the following network and training configurations, which achieve good super-resolution results and avoid overfitting with a low network parameter count. As mentioned, we use a fixed number of 2 \glspl{RSTB} with 4 respective \glspl{STL}. Each \gls{MSA} has 4 attention heads. The feature dimension $C$ is set to 96. We choose a reference model $R_0$ with \gls{DPSWIN} and window size $W=64$ to investigate different processing parts individually. The number of parameters of $R_0$ is 856.065. $R_0$ processes log-compressed rD maps and applies zero-padding for spatial extension.

We use a mini-batch size of 32. The learning rate $lr$ is
\begin{equation}
	lr = 7.5\cdot10^{-4}\cdot0.98477^{epoch}.
\end{equation}
This is a reduction to 10\% $lr$ after 150 epochs. We train with \gls{Adam} \cite{adam}. As standard loss, we use 
the \gls{RMSE} \eqref{eq:trainrmse} between the log-compressed super-resolution rD map and the true, log-compressed high-resolution rD map. Networks are trained for 400 epochs to achieve convergence. For regularization, we apply stochastic depth in the \glspl{STL} \cite{stochastic_depth, swin_arxiv} with survival probability of around $0.91$ in the last range- Doppler \glspl{STL}. In addition, we added Dropout layers with dropout probability $0.05$ after the feature encoder, in the \gls{STL} blocks after the softmax and the final linear dense layer in \gls{MSA}, and after each dense layer in the \gls{MLP}. We also employ gradient clipping to minimize the negative impact of potential outliers on training. CFAR parameters for the CFAR losses are the ones in Fig.~\ref{fig:cfar_masks}, i.e. $P_{FA}=10^{-2}$ and square-shaped guard and reference cell bands of width two and five, respectively.

\subsection{Loss function and evaluation metrics}
\label{sec:metrics}
Our total loss $\mathcal{L}$ is a linear superposition of losses $\mathcal{L}_i$ with weights $\lambda_i$
\begin{equation}
	\mathcal{L}=\sum_{i\in\mathcal{I}}\lambda_{i}\mathcal{L}_i.
\end{equation}
The index $i\in\left\{\mathrm{RMSE},\mathrm{CFAR},\mathrm{wLSD}\right\}$ denotes the loss type. If not specified otherwise $\lambda_i=1$.

As evaluation metrics, we choose \gls{RMSE}
\begin{equation}
	\mathcal{L}_{RMSE} =\sqrt{ \langle\left(\mathbf{X}_{lin}^{hr}-\mathbf{X}_{lin}^{sr}\right)^2\rangle},
\end{equation}
with $\mathbf{X}_{lin}^{hr}$ and $\mathbf{X}_{lin}^{sr}$ being the high/super-resolution rD maps in linear-scale. Additionally, we use \gls{SNR}
\begin{equation}
	\mathcal{L}_{SNR} =10\log_{10}\left( \frac{\langle(\mathbf{X}_{lin}^{hr})^2\rangle}{\langle\left(\mathbf{X}_{lin}^{hr}-\mathbf{X}_{lin}^{sr}\right)^2\rangle}\right),
\end{equation}
a \gls{LSD} \cite{LSD} that's identical to $\mathcal{L}_{RMSE}$ with log-compressed rD maps
\begin{equation}
	\mathcal{L}_{LSD} = \sqrt{\langle\left(\log_{10}\left(\mathbf{X}_{lin}^{hr}\right)-\log_{10}\left(\mathbf{X}_{lin}^{sr}\right)\right)^2\rangle},
	\label{eq:lsd_loss}
\end{equation}
and a bin-wise weighted version wLSD \cite{audio_losses}
\begin{equation}
\mathcal{L}_{wLSD} = \sqrt{\langle\mathbf{W}_{LSD}\circ \left(\log_{10}\left(\mathbf{X}_{lin}^{hr}\right)-\log_{10}\left(\mathbf{X}_{lin}^{sr}\right)\right)^2\rangle}, 
\end{equation}
where we define the weight $\mathbf{W}_{LSD}\in\mathbb{R}^{N_r\times N_D}$ as
\begin{equation}
	\mathbf{W}_{LSD} =  \left(\left(\mathbf{X}_{lin}^{hr}+\mathbf{X}_{lin}^{sr}\right)/2\right)^\beta.
\end{equation}
Opposed to only weighting based on the high-resolution rD map, this symmetric weighting ensures penalization when bins of the super-resolution rD map have large power when they should be weak (target hallucinations). The hyperparameter $\beta$ acts like the $\gamma$ in power-law compression and is set to $\beta = 0.3$ when used for evaluation. We also train with $\mathcal{L}_{wLSD}$ and compare with $\mathcal{L}_{RMSE}$.

Our last evaluation metric is \gls{IoU} (Jaccard index)
\begin{equation}
	\mathrm{IoU} = \frac{\lvert\mathcal{T}_{hr}\cap\mathcal{T}_{sr}\rvert}{\lvert\mathcal{T}_{hr}\cup\mathcal{T}_{sr}\rvert},
\end{equation}
where $\mathcal{T}_{hr}$ is the set of target-plus-interference rD bins of the high-resolution rD map $\mathbf{X}_{lin}^{hr}$ and $\mathcal{T}_{sr}$ of the super-resolution rD map $\mathbf{X}_{lin}^{sr}$. Both sets are computed with identical \gls{CFAR} parameters. We use guard and reference cell bands of width two and five, respectively (as in the CFAR training losses), and evaluate for $P_{FA}\in\left\{10^{-2},10^{-3}\right\}$. We choose high $P_{FA}$ to catch more target bins, especially in clutter-heavy indoor environments such as Fig.~\ref{fig:dataset3}, which make up half of our dataset. This also causes many false-alarms from noise that the network cannot reliably predict, limiting \gls{IoU}.

\section{Results}
\label{sec:res}
\subsection{Attention window and spatial dimension extension}
\begin{figure*}[h]
	\centering
	\subfloat[Super-resolution rD map: pixel-shuffle.]{%
		\includegraphics[width=0.3\textwidth]{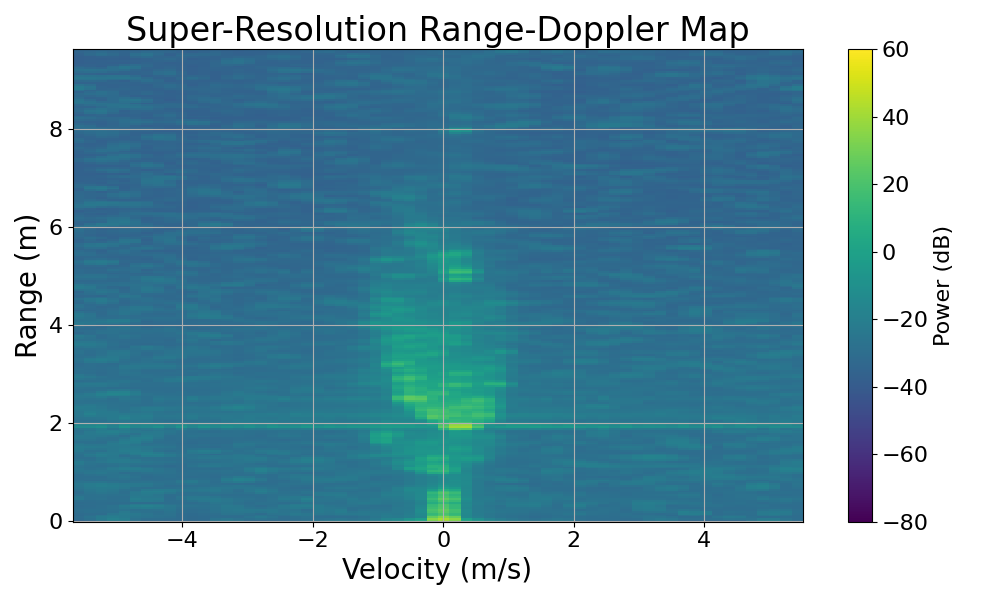}%
	}\hspace{0.03\textwidth}%
	\subfloat[Super-resolution rD map: zero-padding.]{%
		\includegraphics[width=0.3\textwidth]{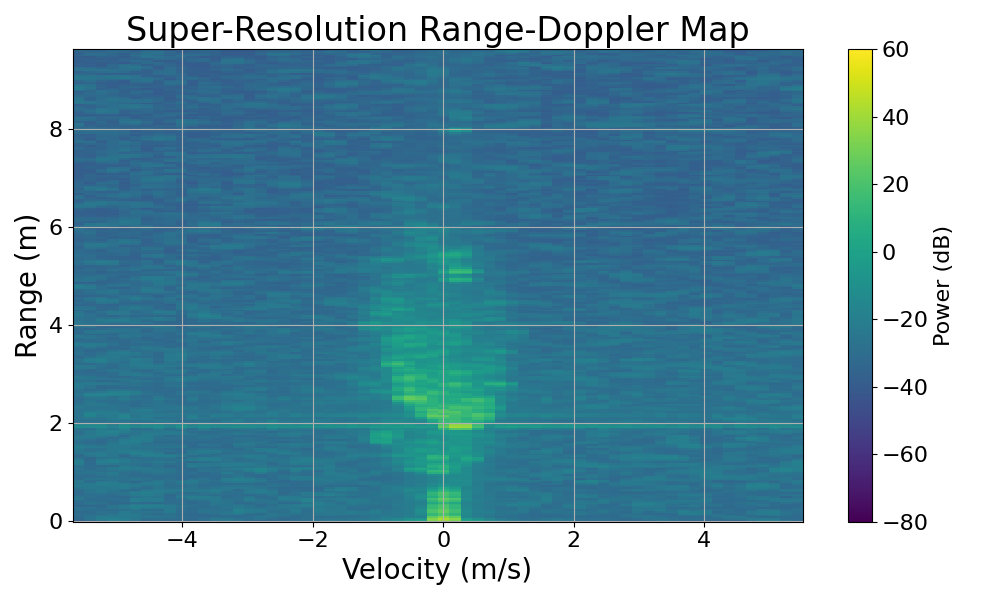}%
	}\hspace{0.03\textwidth}%
	\subfloat[High-resolution rD map.]{%
		\includegraphics[width=0.3\textwidth]{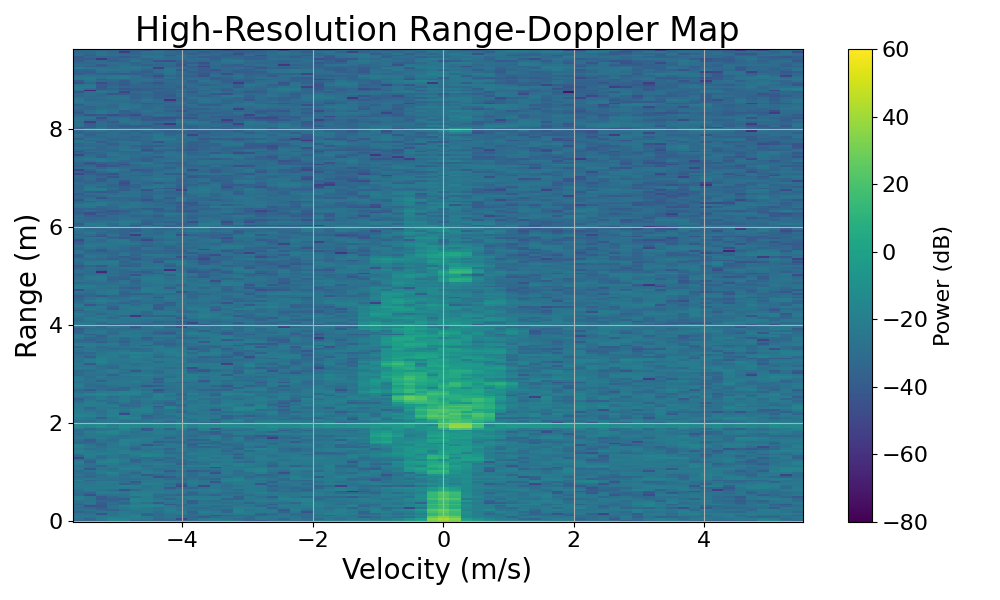}%
	\label{fig:result_gt}
	}
	
	\caption{Effect of spatial dimension extension method on super-resolution rD map. The more expensive zero-padding gives sharper rD maps.}
	\label{fig:results_dim_extension}
\end{figure*}

We evaluate DPSWIN vs. 2D SWIN and zero-padding vs. pixel-shuffle in Tab.~\ref{tab:results_window_up}.
\begin{table}[h]
	\centering
	\caption{Different attention windows, window sizes ($W$), and spatial dimension extension methods.}
	\label{tab:results_window_up}
	\setlength{\tabcolsep}{5pt}
	\begin{tabular}{l c c c c c c c}
		\toprule
		&  & \multicolumn{2}{c}{IoU $\uparrow$} &  &  &  &  \\
		$W$ & 
		& $10^{-2}$ & $10^{-3}$ 
		& RMSE $\downarrow$ 
		& SNR $\uparrow$ 
		& LSD $\downarrow$ 
		& wLSD $\downarrow$ \\
		\midrule		
		\multicolumn{8}{l}{DPSWIN -- zero-padding} \\
		\midrule
		4   &  & 0.4041 & 0.5183 & 0.2454 & 17.91 & 0.2241 & 0.1346 \\
		8   &  & 0.4093 & 0.5295 & 0.2476 & 17.75 & 0.2230 & 0.1338 \\
		16  &  & 0.4137 & 0.5322 & 0.2625 & 17.11 & 0.2218 & 0.1329 \\
		32  &  & 0.4177 & 0.5355 & 0.2535 & 17.48 & 0.2208 & 0.1322 \\
		64  &  & 0.4208 & $\mathbf{0.5409}$ & 0,2467 & 17.81 & 0.2204 & 0.1319 \\
		128 &  & 0.4208 & 0.5406 & 0.2479 & 17.70 & 0.2202 & 0.1317 \\
		256 &  & $\mathbf{0.4220}$ & 0.5396 & 0.2469 & 17.82 & $\mathbf{0.2201}$ & $\mathbf{0.1317}$ \\
		\midrule		
		\multicolumn{8}{l}{DPSWIN -- pixel-shuffle} \\
		\midrule
		8   &  & 0.3352 & 0.4654 & 0.3136 & 15.87 & 0.2491 & 0.1513 \\
		16  &  & 0.3363 & 0.4659 & 0.3137 & 15.91 & 0.2489 & 0.1511 \\
		32  &  & 0.3369 & 0.4647 & 0.3184 & 15.77 & 0.2487 & 0.1509 \\
		64  &  & 0.3377 & 0.4650 & 0.3100 & 16.04 & 0.2486 & 0.1509 \\
		\midrule		
		\multicolumn{8}{l}{2D SWIN -- zero-padding} \\
		\midrule
		4   &  & 0.4062 & 0.5232 & $\mathbf{0.2442}$ & $\mathbf{17.93}$ & 0.2240 & 0.1345 \\
		8   &  & 0.4069 & 0.5239 & 0.2476 & 17.79 & 0.2235 & 0.1341 \\
		16  &  & 0.4105 & 0.5279 & 0.2466 & 17.80 & 0.2227 & 0.1336 \\
		32  &  & 0.4115 & 0.5265 & 0.2445 & 17.91 & 0.2225 & 0.1334 \\
		\midrule	
		\multicolumn{8}{l}{2D SWIN -- pixel-shuffle} \\
		\midrule
		8   &  & 0.3328 & 0.4633 & 0,3119 & 15.95 & 0.2496 & 0.1516 \\
		16  &  & 0.3348 & 0.4646 & 0.3231 & 15.64 & 0.2493 & 0.1514 \\
		\bottomrule
	\end{tabular}
\end{table}
Models using zero-padding for dimension extension outperform the pixel-shuffle versions. We compare the super-resolution output of the DPSWIN model with $W=32$ and pixel-shuffle and the zero-padding model with $W=64$, i.e. windows spanning over the same bandwidth, in Fig.~\ref{fig:results_dim_extension}. Both produce visually pleasing super-resolution rD maps. The version using zero-padding is notably sharper.

The effect of zero-padding is larger than the choice of attention window mechanism. However, it comes at increased computational cost, as the number of rD bins the network has to process increases. Since zero-padding shows effectiveness in super-resolution but adds cost, future work could study if it suffices to do most processing in low-resolution, e.g by downsampling after initial feature extraction. Moreover, higher rD spectrum oversampling  factors, i.e. more zero-padding before the \glspl{FFT}, could be studied. It is also not yet clear if placing pixel-shuffle earlier, or using alternative upsampling methods like transpose convolution \cite{t_conv}, can give improvements.

The best result w.r.t. \gls{LSD} is achieved with large window size and \gls{DPSWIN}. This supports our assumption that large 1D windows are beneficial for our super-resolution task. \gls{DPSWIN} \gls{LSD} metrics tend to become better with larger window size and small windows already work very well. We assumed increasing receptive field through stacking \glspl{STL} in combination with small windows and \gls{DPSWIN} might be sufficient (and computationally cheaper), which also seems to hold.  

Models using 2D \gls{SWIN} are overall worse in LSD, even though they use larger windows with $W^2$ opposed to $W$ tokens in \gls{DPSWIN}. 2D SWIN behaves similar as DPSWIN, but the LSD gain with larger windows is smaller. The reason is probably that 
rD maps have dominant 2D structures directly around targets,
not at far distances. Larger 2D windows likely contain too many uncorrelated and noisy rD bins that complicate learning. While 2D \gls{SWIN} is worse in LSD, wLSD, and IoU, the best model in  \gls{RMSE} and \gls{SNR} uses 2D windows with $W=4$. As noted, both \gls{RMSE} and \gls{SNR} mainly reflect the reconstruction of the strongest rD bins, as bins with lower power have little effect on them, and we believe that the LSD metrics are more representative of super-resolution performance.

\subsection{CFAR losses}
The \gls{CFAR} losses should align the CFAR behavior of high- and super-resolution rD maps. Our goal was to test if we can reduce blurring and improve super-resolution signal reconstruction and target detection metrics.

In preliminary experiments, we ruled out further evaluation of $t^1$ and $t^2$, as $t^3$ gave better results. We combine $\mathcal{L}_{RMSE}$ with $\lambda_{RMSE}=0.5$ and $\mathcal{L}_{CFAR}$. All models are trained as DPSWIN models with $W=64$ and zero-padding and we use the respective model from Tab.~\ref{tab:results_window_up} as reference model $R_0$ trained purely on $\mathcal{L}_{RMSE}$ with $\lambda_{RMSE}=1.0$. The CFAR loss results are in Tab.~\ref{tab:results_cfar_mixed}.
\begin{table}[h]
	\centering
	\caption{CFAR losses in combination with $L_{RMSE}$. The MSE versions where $\xi$ is "--" skip the sigmoid. }
	\label{tab:results_cfar_mixed}
	\setlength{\tabcolsep}{2pt}
	\begin{tabular}{l c c c c c c c c}
		\toprule
		& & & \multicolumn{2}{c}{IoU $\uparrow$} & & & & \\
		Loss & $\xi$ & $\lambda_{\text{CFAR}}$ 
		& $10^{-2}$ & $10^{-3}$ 
		& RMSE $\downarrow$ 
		& SNR $\uparrow$ 
		& LSD $\downarrow$ 
		& wLSD $\downarrow$ \\
		\midrule
		BCE & 0.1 & 0.5   & 0.4176 & 0.5323 & 0.2443 & 17.81 & 0.2207 & 0.1322 \\
		BCE & 1   & 0.5   & $\mathbf{0.4258}$ & 0.5553 & 0.2262 & 18.65 & 0.2204 & $\mathbf{0.1318}$ \\
		BCE & 1   & 5     & 0.4231 & 0.5594 & 0.2392 & 18.13 & 0.2209 & 0.1322 \\
		BCE & 1   & 50    & 0.4214 & 0.5611 & 0.2237 & 18.65 & 0.2235 & 0.1339 \\
		KLD & 0.1 & 0.5   & 0.4184 & 0.5358 & 0.2368 & 18.15 & 0.2207 & 0.1321 \\
		KLD & 1   & 0.5   & 0.4250 & 0.5621 & 0.2956 & 16.09 & 0.2207 & 0.1321 \\
		KLD & 1   & 5     & 0.4239 & $\mathbf{0.5662}$ & 0.4498 & 12.17 & 0.2215 & 0.1329 \\
		KLD & 1   & 50    & 0.4238 & 0.5647 & 0.5208 & 10.86 & 0.2221 & 0.1335 \\
		MSE & 1   & 0.5   & 0.4245 & 0.5505 & 0.2437 & 17.84 & $\mathbf{0.2203}$ & $\mathbf{0.1318}$ \\
		MSE & 1   & 5     & 0.4222 & 0.5612 & $\mathbf{0.2201}$ & $\mathbf{18.82}$ & 0.2206 & 0.1319 \\
		MSE & 1   & 50    & 0.4151 & 0.5593 & 0.2272 & 18.59 & 0.2228 & 0.1334 \\
		MSE & --  & 0.005 & 0.4153 & 0.5232 & 0.2299 & 18.44 & 0.2204 & 0.1319 \\
		MSE & --  & 0.05  & 0.3792 & 0.4266 & 0.2438 & 17.78 & 0.2213 & 0.1326 \\
		MSE & --  & 0.5   & 0.3642 & 0.3696 & 0.3902 & 13.46 & 0.2233 & 0.1343 \\
		\midrule
		$R_0$ & -- & -- & 0.4208 & 0.5409 & 0.2467 & 17.81 & 0.2204 & 0.1319 \\
		\bottomrule
	\end{tabular}
\end{table}
The models trained on KLD with large $\lambda_{CFAR}$ achieve the best IoU with $P_{FA}=10^{-3}$, but this comes with a huge drop in RMSE and SNR compared with $R_0$. Models trained on MSE where the sigmoid was skipped, i.e. $m_{k,l}=t^3_{k,l}$, give no notable gain over $R_0$, where setting $\lambda_{CFAR}$ larger only degrades the performance. The MSE models with the sigmoid, however, overall outperform $R_0$. The MSE model with $\lambda_{CFAR}=0.5$ beats $R_0$ in every objective. The one with $\lambda_{CFAR}=5.0$ has the largest advantage over the $R_0$ in RMSE and SNR with a minor increase in LSD. The BCE losses work similarly well, where the $\lambda_{CFAR}=0.5$ model has the best IoU for $P_{FA}=10^{-2}$ and is better than $R_0$ in all objectives except LSD, where it's identical. As with the MSE losses, the BCE losses show that increasing $\lambda_{CFAR}$ can possibly improve IoU with $P_{FA}=10^{-3}$, RMSE, and SNR, but at the cost of worse LSD and wLSD. The steeper sigmoids with $\xi=1$ produce better results than with $\xi=0.1$. Moreover, we expect that a wider hyperparameter search for the CFAR parameters, stacking different CFAR masks, or integrating class balancing \cite{cfar_loss,cfar_loss_diffusion_augmentation}, could bring further gains.

Even though the metrics improve, we could not notice a visible reduction in blurring, which is likely too small to see. An alternative technique to mitigate blurring in future work is the integration of spectral layers as in \cite{swinfir}, i.e. layers including Fourier, wavelet, cosine, or similar spectral transforms, to facilitate processing of high-frequency rD map components. Similarly, one might add losses defined in the spectrum of the compressed super-resolution rD map, i.e. after transformation of the compressed rD map with another 2D FFT. This allows direct penalization of visual high-frequency errors in the super-resolution rD maps.

\subsection{Weighted LSD loss}
We alternatively train with $\mathcal{L}_{wLSD}$ instead of $\mathcal{L}_{RMSE}$ and show the results in Tab.~\ref{tab:results_wlds}.

\begin{table}[h]
	\centering
	\caption{Training with $\mathcal{L}_{wLSD}$ instead of $\mathcal{L}_{RMSE}$ ($R_0$).}
	\label{tab:results_wlds}
	\setlength{\tabcolsep}{2pt}
	\begin{tabular}{l c c c c c c c c}
		\toprule
		& & & \multicolumn{2}{c}{IoU $\uparrow$} & & & & \\
		Loss & $\beta$ & $\lambda_{wLSD}$ 
		& $10^{-2}$ & $10^{-3}$ 
		& RMSE $\downarrow$ 
		& SNR $\uparrow$ 
		& LSD $\downarrow$ 
		& wLSD $\downarrow$ \\
		\midrule
		wLSD & 0.2 & 0.1 & 0.4169 & 0.5355 & 0.2212 & 18.82 & 0.2217 & 0.1327 \\
		wLSD & 0.2 & 1   & 0.4206 & 0.5403 & 0.2172 & 18.99 & 0.2207 & 0.1320 \\
		wLSD & 0.3 & 0.1 & 0.4158 & 0.5338 & 0.2145 & 19.14 & 0.2222 & 0.1330 \\
		wLSD & 0.3 & 1   & $\mathbf{0.4212}$ & 0.5426 & 0.2161 & 18.99 & 0.2209 & 0.1321 \\
		wLSD & 0.5 & 0.1 & 0.4159 & 0.5387 & 0.2077 & 19.43 & 0.2230 & 0.1334 \\
		wLSD & 0.5 & 1   & 0.4206 & $\mathbf{0.5446}$ & $\mathbf{0.2015}$ & $\mathbf{19.71}$ & 0.2219 & 0.1327 \\
		\midrule
		$R_0$ & -- & -- & 0.4208 & 0.5409 & 0.2467 & 17.81 & $\mathbf{0.2204}$ & $\mathbf{0.1319}$ \\
		\bottomrule
	\end{tabular}
\end{table}

The additional weighting with compressed signal power has a small effect on the \gls{LSD} metrics, which increases with lower compression (larger $\beta$). Larger $\beta$ will cause more pronounced down-weighting of low-power bins. The effect on the other metrics follows our expectations. Weighting the log-scale rD bins causes a similar shift in metrics as with the CFAR losses. Adding CFAR losses is more effective in increasing \gls{IoU}, thus likely in improving downstream object detection and segmentation, but wLSD with lower weight compression (higher $\beta$) gives the best RMSE and SNR.

\subsection{Compression}
Last, we investigate the effect of magnitude compression in the rD map in Tab.~\ref{tab:comp_norm}. S+ are models that include the output softplus. 

 Remember that the $\mathcal{L}_{RMSE}$ loss is based on the network input/output compressed representation. For power-law compression, we set $\gamma=0.3$. In Fig.~\ref{fig:compression}, we could see that for our scenarios, power-law compression allows for a likely better trade-off in target structure vs. interference level and $\gamma=0.3$ seems like a good choice. 

\begin{table}[h]
	\centering
	\caption{Effect of rD bin magnitude compression.}
	\label{tab:comp_norm}
	\setlength{\tabcolsep}{2pt}
	\begin{tabular}{l c c c c c c c c}
		\toprule
		& & & & \multicolumn{2}{c}{IoU $\uparrow$} & $\downarrow$ & $\uparrow$ & $\downarrow$\\
		comp & S+ 
		& $\lambda_{\text{RMSE}}$ & $\lambda_{\text{LSD}}$
		& $10^{-2}$ & $10^{-3}$ 
		& RMSE
		& SNR
		& LSD\\
		\midrule
		lin  & False & 0.1 & -- & 0.3363 & 0.5053 & 0.2130 & 19.33 & 0.4872 \\
		lin  & False & 1   & -- & 0.3388 & 0.5047 & 0.2127 & 19.32 & 0.4638 \\
		lin  & True  & 0.1 & -- & 0.3484 & 0.4930 & 0.2135 & 19.31 & 0.2718 \\
		lin  & True  & --  & 1  & 0.4204 & 0.5415 & 0.6381 &  9.11 & 0.2205 \\
		pow & False & 1   & -- & 0.4182 & $\mathbf{0.5427}$ & 0.2005 & 19.80 & 0.2236 \\
		pow & True  & 1   & -- & 0.4174 & 0.5385 & $\mathbf{0.2001}$ & $\mathbf{19.81}$ & 0.2236 \\
		pow & True  & --  & 1  & 0.4159 & 0.5313 & 0.2522 & 17.71 & 0.2213 \\
		\midrule
		$R_0$ (log)& False & 1   & -- & $\mathbf{0.4208}$ & 0.5409 & 0.2467 & 17.81 & $\mathbf{0.2204}$ \\
		\bottomrule
	\end{tabular}
\end{table}
\begin{figure*}[h!]
	\centering
	\subfloat[Linear-scale.]{%
		\includegraphics[width=0.3\textwidth]{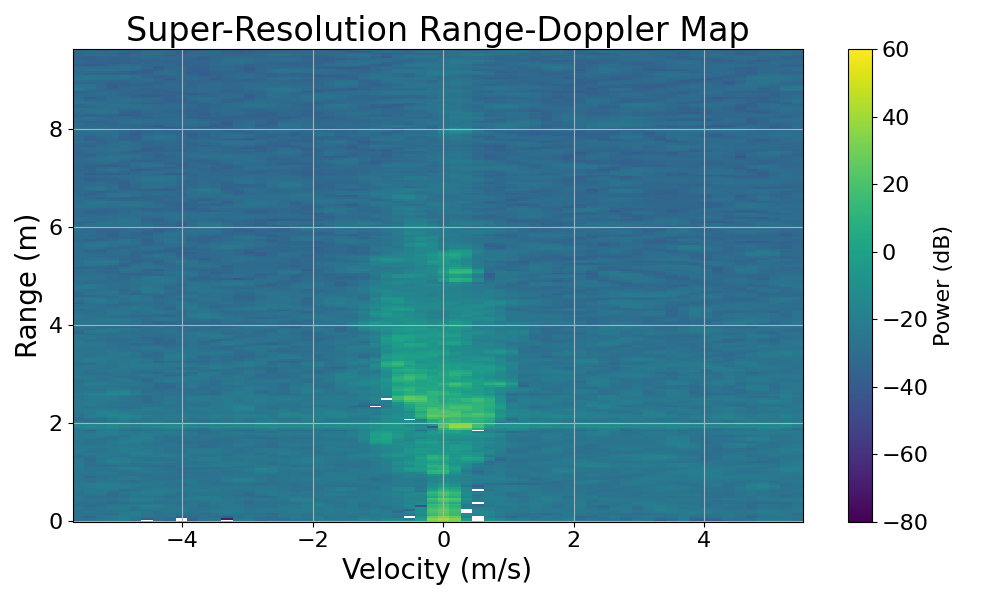}%
	\label{fig:result_linscale}
	}\hspace{0.03\textwidth}%
	\subfloat[Power-law.]{%
		\includegraphics[width=0.3\textwidth]{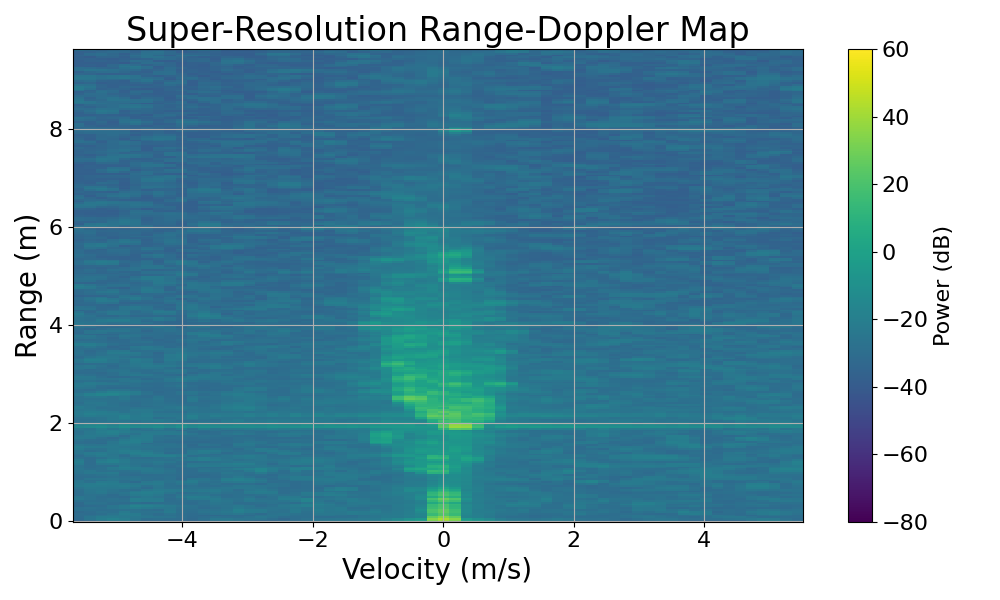}%
	}\hspace{0.03\textwidth}%
	\subfloat[Log-scale]{%
		\includegraphics[width=0.3\textwidth]{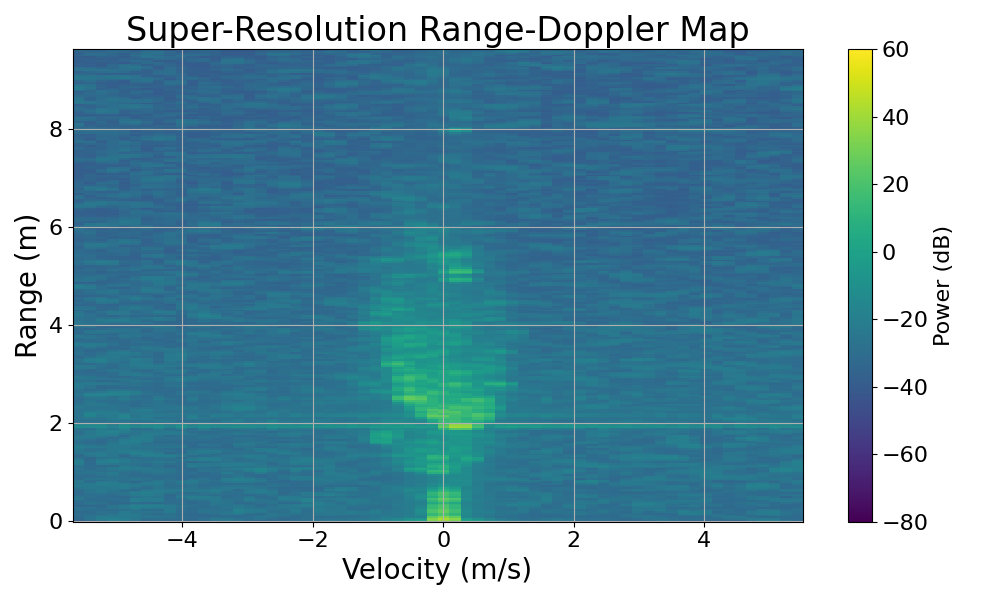}%
	}
	\caption{Effect of compression method on super-resolution rD map.}
	\label{fig:results_compression}
\end{figure*}
No compression (lin) and $\mathcal{L}_{RMSE}$ gives better RMSE and SNR than $R_0$. This is expected, as in this case $\mathcal{L}_{RMSE}$ is computed based on the linear-scale rD maps. However, IoU is worse and LSD is very bad. When using a linear-scale input and the log-compression loss $\mathcal{L}_{LSD}$, interestingly good IoU and \gls{LSD} are achieved. However, this model has issues in RMSE and SNR. $R_0$ with log-compressed inputs trained on the log-compression loss unsurprisingly produces the best LSD, and also the best IoU for $P_{FA}=10^{-2}$. Using power-law compressed inputs with $\mathcal{L}_{LSD}$ works better than the version with linear-scale inputs and $\mathcal{L}_{LSD}$, but is consistently worse than $R_0$. However, the combination of power-law compressed inputs and power-law compression loss gives similar IoU as $R_0$, with a slight increase in LSD, but a notable improvement in RMSE and SNR. 

Fig.~\ref{fig:results_compression} compares the outputs of the $\mathcal{L}_{RMSE}$ models  without the softplus, which did not improve results. The model using linear rD maps struggles with reconstruction of bins other than the main scatterer. This is in accordance with our expectation, as those bins are badly visible and have too little loss contribution in linear scale. The difference between power-law and log-scale is less pronounced. The smoothing with power-law compression is slightly lower, which explains the better RMSE and SNR, and the worse LSD which profits from over-smoothing noise.

Overall, power-law compression gives the most well rounded results for our dataset with a small advantage over log-compression and a large one over no compression.

\section{Conclusion}
\label{sec:con}
We investigated deep learning-based super-resolution of \gls{rD} maps from a \gls{CS} radar. Opposed to existing literature, we studied real environments such as subway stations and shopping streets. We introduced an efficient SWIN attention mechanism \gls{DPSWIN}, which has shown strong results at low processing cost. We applied and compared different CFAR losses, in addition to the standard RMSE loss, to the super-resolution task, and demonstrated that integrating such losses improves RMSE, SNR, and \gls{IoU}. All design choices in our super-resolution framework like magnitude compression in the rD map, zero-padding before FFTs, 1D attention windows etc., have demonstrated clearly their advantages.

Our work has several limitations. The supervised training, while simple to realize, has multiple issues that limit the direct application of the rD super-resolution method for downstream tasks in practice. This includes possible range cell migration effects in the high-resolution signals and incomplete low-resolution chirps after dropping high-resolution fast-time samples. We have so far only studied super-resolution methodology without downstream task like target localization. Further, while we did initial benchmarks against simple CNNs and U-Nets, benchmarks against similar models from similar tasks, e.g. segmentation models from the automotive radar domain, and on their datasets, would be of high interest.

In future work, solving the open (low-res, high-res) tuple generation problems, adding more benchmarks w.r.t. other models, datasets, and downstream tasks, and studying the limits of upsampling factors, the number of stacked rD maps in the input etc., are most important. We believe the low-cost \gls{DPSWIN} attention, the CFAR losses, and the focus on representations (zero-padding before FFTs etc.), translate well to many other radar applications.

\section{Acknowledgements}
The authors acknowledge that ChatGPT (free version) was used to test ideas in code, for additional literature research, and to improve text.

\bibliographystyle{IEEEtran}
\bibliography{ref}

@article{layernorm,
	title={Layer normalization},
	author={Ba, Jimmy Lei and Kiros, Jamie Ryan and Hinton, Geoffrey E},
	journal={arXiv preprint arXiv:1607.06450},
	year={2016}
}

@book{stft,
	title={Spectral Audio Signal Processing},
	author={Smith, J.O. and Stanford University. Center for Computer Research in Music and Acoustics and Stanford University. Department of Music},
	isbn={9780974560731},
	year={2011},
	publisher={W3K}
}

@misc{tensorflow2015-whitepaper,
	title={ {TensorFlow}: Large-Scale Machine Learning on Heterogeneous Systems},
	url={https://www.tensorflow.org/},
	author={
	Mart\'{i}n~Abadi and
	others},
	year={2015},
}

@article{transformer,
	title={Attention is all you need},
	author={Vaswani, Ashish and Shazeer, Noam and Parmar, Niki and Uszkoreit, Jakob and Jones, Llion and Gomez, Aidan N and Kaiser, {\L}ukasz and Polosukhin, Illia},
	journal={Advances in neural information processing systems},
	volume={30},
	year={2017}
}

@book{dl_goodfellow,
	title={Deep Learning},
	author={Ian Goodfellow and Yoshua Bengio and Aaron Courville},
	publisher={MIT Press},
	note={\url{http://www.deeplearningbook.org}},
	year={2016}
}

@article{t_conv,
	title={A guide to convolution arithmetic for deep learning},
	author={Dumoulin, Vincent and Visin, Francesco},
	journal={arXiv preprint arXiv:1603.07285},
	year={2016}
}

@techreport{fft_windows,
	author      = {Heinzel, Gerhard and R{\"u}diger, Albrecht and Schilling, Roland},
	title       = {Spectrum and spectral density estimation by the Discrete {Fourier} transform ({DFT}), including a comprehensive list of window functions and some new flat-top windows},
	institution = {Max-Planck-Institut f{\"u}r Gravitationsphysik, Hannover},
	year        = {2002}
}

@INPROCEEDINGS{cfar_loss,
	author={Oswald, Christian and Toth, Mate and Meissner, Paul and Pernkopf, Franz},
	booktitle={2023 IEEE International Radar Conference (RADAR)}, 
	title={End-to-End Training of Neural Networks for Automotive Radar Interference Mitigation}, 
	year={2023},
	volume={},
	number={},
	pages={1-6},
	doi={10.1109/RADAR54928.2023.10371003}}

@InProceedings{swin_unet,
	author="Cao, Hu
	and Wang, Yueyue
	and Chen, Joy
	and Jiang, Dongsheng
	and Zhang, Xiaopeng
	and Tian, Qi
	and Wang, Manning",
	editor="Karlinsky, Leonid
	and Michaeli, Tomer
	and Nishino, Ko",
	title="Swin-{Unet}: Unet-Like Pure Transformer for Medical Image Segmentation",
	booktitle="Computer Vision -- ECCV 2022 Workshops",
	year="2023",
	publisher="Springer Nature Switzerland",
	address="Cham",
	pages="205--218",
	isbn="978-3-031-25066-8"
}

@INPROCEEDINGS{rD_diffusion,
	author={Qosja, Denisa and Barth, Kilian and Wagner, Simon},
	booktitle={2025 26th International Radar Symposium (IRS)}, 
	title={Enhancing {Fourier}-Based {Doppler} Resolution with Diffusion Models}, 
	year={2025},
	volume={},
	number={},
	pages={1-6},
	doi={10.23919/IRS64527.2025.11046167}}

@INPROCEEDINGS{pixel_shuffle,
	author={Shi, Wenzhe and Caballero, Jose and Huszár, Ferenc and Totz, Johannes and Aitken, Andrew P. and Bishop, Rob and Rueckert, Daniel and Wang, Zehan},
	booktitle={2016 IEEE Conference on Computer Vision and Pattern Recognition (CVPR)}, 
	title={Real-Time Single Image and Video Super-Resolution Using an Efficient Sub-Pixel Convolutional Neural Network}, 
	year={2016},
	volume={},
	number={},
	pages={1874-1883},
	doi={10.1109/CVPR.2016.207}}

@INPROCEEDINGS{vit_radar_gait_recognition,
	author={Chen, Shiliang and He, Wentao and Ren, Jianfeng and Jiang, Xudong},
	booktitle={ICASSP 2022 - 2022 IEEE International Conference on Acoustics, Speech and Signal Processing (ICASSP)}, 
	title={Attention-Based Dual-Stream Vision Transformer for Radar Gait Recognition}, 
	year={2022},
	volume={},
	number={},
	pages={3668-3672},
	doi={10.1109/ICASSP43922.2022.9746565}}

@ARTICLE{dp_radar_people_counting,
	author={Jiang, Xikang and Guo, Jiahang and Rao, Chong and Zhang, Lin and Li, Lei},
	journal={IEEE Transactions on Radar Systems}, 
	title={{UP-TIFA}: {UWB} Radar-Based People Counting via Time–Frequency Attention Neural Network}, 
	year={2025},
	volume={3},
	number={},
	pages={1233-1242},
	doi={10.1109/TRS.2025.3605232}}

@ARTICLE{radar_swin_3D_object_detection,
	author={Jiang, Tiezhen and Zhuang, Long and An, Qi and Wang, Jianhua and Xiao, Kai and Wang, Anqi},
	journal={IEEE Transactions on Instrumentation and Measurement}, 
	title={{T-RODNet}: Transformer for Vehicular Millimeter-Wave Radar Object Detection}, 
	year={2023},
	volume={72},
	number={},
	pages={1-12},
	doi={10.1109/TIM.2022.3229703}}

@book{richards_modern_radar,
	author = {Mark A. Richards  and James A. Scheer  and William A. Holm },
	title = {Principles of Modern Radar: Basic principles},
	publisher = {The Institution of Engineering and Technology},
	year = {2010},
	doi = {10.1049/SBRA021E},
	edition   = {},
	eprint = {https://digital-library.theiet.org/doi/pdf/10.1049/SBRA021E}
}

@INPROCEEDINGS{irs22_radar_separation,
	author={Hinderer, Sven},
	booktitle={2022 23rd International Radar Symposium (IRS)}, 
	title={Blind Source Separation of Radar Signals in Time Domain Using Deep Learning}, 
	year={2022},
	volume={},
	number={},
	pages={486-491},
	doi={10.23919/IRS54158.2022.9904990}}

@INPROCEEDINGS{radar_superres_karim_sherif,
	author={Armanious, Karim and Abdulatif, Sherif and Aziz, Fady and Schneider, Urs and Yang, Bin},
	booktitle={2019 27th European Signal Processing Conference (EUSIPCO)}, 
	title={An Adversarial Super-Resolution Remedy for Radar Design Trade-offs}, 
	year={2019},
	volume={},
	number={},
	pages={1-5},
	doi={10.23919/EUSIPCO.2019.8902510}}

@INPROCEEDINGS{SwinIR,
	author={Liang, Jingyun and Cao, Jiezhang and Sun, Guolei and Zhang, Kai and Van Gool, Luc and Timofte, Radu},
	booktitle={2021 IEEE/CVF International Conference on Computer Vision Workshops (ICCVW)}, 
	title={{SwinIR}: Image Restoration Using {Swin} Transformer}, 
	year={2021},
	volume={},
	number={},
	pages={1833-1844},
	doi={10.1109/ICCVW54120.2021.00210}}

@INPROCEEDINGS{SepFormer,
	author={Subakan, Cem and Ravanelli, Mirco and Cornell, Samuele and Bronzi, Mirko and Zhong, Jianyuan},
	booktitle={ICASSP 2021 - 2021 IEEE International Conference on Acoustics, Speech and Signal Processing (ICASSP)}, 
	title={Attention Is All You Need In Speech Separation}, 
	year={2021},
	volume={},
	number={},
	pages={21-25},
	doi={10.1109/ICASSP39728.2021.9413901}}

@inproceedings{dptnet,
	title     = {Dual-Path Transformer Network: Direct Context-Aware Modeling for End-to-End Monaural Speech Separation},
	author    = {Jingjing Chen and Qirong Mao and Dong Liu},
	year      = {2020},
	booktitle = {{Interspeech 2020}},
	pages     = {2642--2646},
	doi       = {10.21437/Interspeech.2020-2205},
	issn      = {2958-1796},
}

@article{axial_attention,
	author       = {Jonathan Ho and
	Nal Kalchbrenner and
	Dirk Weissenborn and
	Tim Salimans},
	title        = {Axial Attention in Multidimensional Transformers},
	journal      = {CoRR},
	volume       = {abs/1912.12180},
	year         = {2019},
	eprinttype    = {arXiv},
	eprint       = {1912.12180},
}

@Article{superres_sar,
	AUTHOR = {Kong, Yingying and Liu, Si},
	TITLE = {{DMSC-GAN}: A {c-GAN}-Based Framework for Super-Resolution Reconstruction of {SAR} Images},
	JOURNAL = {Remote Sensing},
	VOLUME = {16},
	YEAR = {2024},
	NUMBER = {1},
	ARTICLE-NUMBER = {50},
	ISSN = {2072-4292},
	DOI = {10.3390/rs16010050}
}

@ARTICLE{rD_superres_unet,
	author={Jeong, Taewon and Lee, Seongwook},
	journal={IEEE Access}, 
	title={Resource-Efficient Range-{Doppler} Map Generation Using Deep Learning Network for Automotive Radar Systems}, 
	year={2023},
	volume={11},
	number={},
	pages={55965-55977},
	doi={10.1109/ACCESS.2023.3282688}}

@article {ppi_unet,
	author = "Andrew Geiss and Joseph C. Hardin",
	title = "Radar Super Resolution Using a Deep Convolutional Neural Network",
	journal = "Journal of Atmospheric and Oceanic Technology",
	year = "2020",
	publisher = "American Meteorological Society",
	address = "Boston MA, USA",
	volume = "37",
	number = "12",
	doi = "10.1175/JTECH-D-20-0074.1",
	pages=      "2197 - 2207",
}

@INPROCEEDINGS{az_superres,
	author={Li, Yu-Jhe and Hunt, Shawn and Park, Jinhyung and O'Toole, Matthew and Kitani, Kris},
	booktitle={2023 IEEE/CVF Conference on Computer Vision and Pattern Recognition (CVPR)}, 
	title={Azimuth Super-Resolution for {FMCW} Radar in Autonomous Driving}, 
	year={2023},
	volume={},
	number={},
	pages={17504-17513},
	doi={10.1109/CVPR52729.2023.01679}}

@INPROCEEDINGS{4d_superres,
	author={Prabhakara, Akarsh and Jin, Tao and Das, Arnav and Bhatt, Gantavya and Kumari, Lilly and Soltanaghai, Elahe and Bilmes, Jeff and Kumar, Swarun and Rowe, Anthony},
	booktitle={2023 IEEE International Conference on Robotics and Automation (ICRA)}, 
	title={High Resolution Point Clouds from mmWave Radar}, 
	year={2023},
	volume={},
	number={},
	pages={4135-4142},
	doi={10.1109/ICRA48891.2023.10161429}}

@INPROCEEDINGS{trans_radar,
	author={Dalbah, Yahia and Lahoud, Jean and Cholakkal, Hisham},
	booktitle={2024 IEEE/CVF Winter Conference on Applications of Computer Vision (WACV)}, 
	title={{TransRadar}: Adaptive-Directional Transformer for Real-Time Multi-View Radar Semantic Segmentation}, 
	year={2024},
	volume={},
	number={},
	pages={352-361},
	doi={10.1109/WACV57701.2024.00042}}

@INPROCEEDINGS{neighborhood_attn,
	author={Hassani, Ali and Walton, Steven and Li, Jiachen and Li, Shen and Shi, Humphrey},
	booktitle={2023 IEEE/CVF Conference on Computer Vision and Pattern Recognition (CVPR)}, 
	title={Neighborhood Attention Transformer}, 
	year={2023},
	volume={},
	number={},
	pages={6185-6194},
	doi={10.1109/CVPR52729.2023.00599}}

@ARTICLE{rD_upscale_complex,
	author={Ding, Minhao and Ding, Yipeng and Lv, Ping and Tang, Bowen and Liu, Runjin},
	journal={IEEE Geoscience and Remote Sensing Letters}, 
	title={{HRRD-Net}: An End-to-End High-Resolution Range-{Doppler} Spectrum Estimation Algorithm}, 
	year={2025},
	volume={22},
	number={},
	pages={1-4},
	doi={10.1109/LGRS.2024.3521023}}

@INPROCEEDINGS{lidar_upsampling_rect_swin,
	author={Yang, Bin and Pfreundschuh, Patrick and Siegwart, Roland and Hutter, Marco and Moghadam, Peyman and Patil, Vaishakh},
	booktitle={2024 IEEE/CVF Conference on Computer Vision and Pattern Recognition (CVPR)}, 
	title={{TULIP}: Transformer for Upsampling of {LiDAR} Point Clouds}, 
	year={2024},
	volume={},
	number={},
	pages={15354-15364},
	doi={10.1109/CVPR52733.2024.01454}}

@PhdThesis{phd_chenming,
	author = 	{Jiang, Chenming},
	title = 	{Generative models for interference mitigation and range-{Doppler} super-resolution for automotive radar},
	year = 	{2025},
	month = 	{Jun},
	school = {Universit{\"a}t Stuttgart},
	address = 	{Stuttgart},
	doi = 	{},
	url = 	{},
	language = 	{en}
}

@INPROCEEDINGS{swinv2,
	author={Liu, Ze and Hu, Han and Lin, Yutong and Yao, Zhuliang and Xie, Zhenda and Wei, Yixuan and Ning, Jia and Cao, Yue and Zhang, Zheng and Dong, Li and Wei, Furu and Guo, Baining},
	booktitle={2022 IEEE/CVF Conference on Computer Vision and Pattern Recognition (CVPR)}, 
	title={{Swin Transformer V2}: Scaling Up Capacity and Resolution}, 
	year={2022},
	volume={},
	number={},
	pages={11999-12009},
	doi={10.1109/CVPR52688.2022.01170}}

@misc{swinfir,
	title={{SwinFIR}: Revisiting the {SwinIR} with Fast {Fourier} Convolution and Improved Training for Image Super-Resolution}, 
	author={Dafeng Zhang and Feiyu Huang and Shizhuo Liu and Xiaobing Wang and Zhezhu Jin},
	year={2023},
	eprint={2208.11247},
	 url={https://arxiv.org/abs/2208.11247}, 
	archivePrefix={arXiv},
	primaryClass={cs.CV},
}

@INPROCEEDINGS{swin,
	author={Liu, Ze and Lin, Yutong and Cao, Yue and Hu, Han and Wei, Yixuan and Zhang, Zheng and Lin, Stephen and Guo, Baining},
	booktitle={2021 IEEE/CVF International Conference on Computer Vision (ICCV)}, 
	title={Swin Transformer: Hierarchical Vision Transformer using Shifted Windows}, 
	year={2021},
	volume={},
	number={},
	pages={9992-10002},
	doi={10.1109/ICCV48922.2021.00986}}

@InProceedings{clip,
	title = 	 {Learning Transferable Visual Models From Natural Language Supervision},
	author =       {Radford, Alec and Kim, Jong Wook and Hallacy, Chris and Ramesh, Aditya and Goh, Gabriel and Agarwal, Sandhini and Sastry, Girish and Askell, Amanda and Mishkin, Pamela and Clark, Jack and Krueger, Gretchen and Sutskever, Ilya},
	booktitle = 	 {Proceedings of the 38th International Conference on Machine Learning},
	pages = 	 {8748--8763},
	year = 	 {2021},
	editor = 	 {Meila, Marina and Zhang, Tong},
	volume = 	 {139},
	series = 	 {Proceedings of Machine Learning Research},
	month = 	 {18--24 Jul},
	publisher =    {PMLR}
}

@inproceedings{vit,
	author       = {Alexey Dosovitskiy and
	Lucas Beyer and
	Alexander Kolesnikov and
	Dirk Weissenborn and
	Xiaohua Zhai and
	Thomas Unterthiner and
	Mostafa Dehghani and
	Matthias Minderer and
	Georg Heigold and
	Sylvain Gelly and
	Jakob Uszkoreit and
	Neil Houlsby},
	title        = {An Image is Worth 16x16 Words: Transformers for Image Recognition
	at Scale},
	booktitle    = {9th International Conference on Learning Representations, {ICLR} 2021,
	Virtual Event, Austria, May 3-7, 2021},
	publisher    = {OpenReview.net},
	year         = {2021},
	bibsource    = {dblp computer science bibliography, https://dblp.org}}

@manual{infineon2023bgt60tr13c_datasheet,
	title        = "{BGT60TR13C Datasheet V2.4.9}",
	author       = "{Infineon Technologies AG}",
	year         = 2023,
	month        = Nov,
	url	= {https://www.infineon.com/cms/en/product/sensor/radar-sensors/radar-sensors-for-iot/60ghz-radar/bgt60tr13c/},
	note         = "{Accessed}: Feb. 28, 2026",
	institution  = "Infineon Technologies",
	address      = "Munich, Germany"
}

@inproceedings{diffusion,
	author = {Ho, Jonathan and Jain, Ajay and Abbeel, Pieter},
	title = {Denoising diffusion probabilistic models},
	year = {2020},
	isbn = {9781713829546},
	publisher = {Curran Associates Inc.},
	address = {Red Hook, NY, USA},
	booktitle = {Proceedings of the 34th International Conference on Neural Information Processing Systems},
	articleno = {574},
	numpages = {12},
	location = {Vancouver, BC, Canada},
	series = {NIPS '20}
}

@INPROCEEDINGS{irs25,
	author={Hinderer, Sven and Yin, Zheming and Papanikolaou, Athanasios and Hesselbarth, Jan and Yang, Bin},
	booktitle={2025 26th International Radar Symposium (IRS)}, 
	title={Hybrid Baseband Simulation for Single-Channel Radar-Based Indoor Localization System}, 
	year={2025},
	volume={},
	number={},
	pages={1-8},
	doi={10.23919/IRS64527.2025.11046115}}

@InProceedings{unet,
	author       = "O. Ronneberger and P. Fischer and T. Brox",
	title        = "{U-Net}: Convolutional Networks for Biomedical Image Segmentation",
	booktitle    = "Medical Image Computing and Computer-Assisted Intervention (MICCAI)",
	series       = "LNCS",
	volume       = "9351",
	pages        = "234--241",
	year         = "2015",
	publisher    = "Springer",

}

@article{cfarnet,
	title = {{CFARnet}: Deep learning for target detection with constant false alarm rate},
	journal = {Signal Processing},
	volume = {223},
	pages = {109543},
	year = {2024},
	issn = {0165-1684},
	doi = {https://doi.org/10.1016/j.sigpro.2024.109543},
	author = {Tzvi Diskin and Yiftach Beer and Uri Okun and Ami Wiesel}
}

@inproceedings{bgt_rl_tracking,
	author       = {Julius Ott and
	Lorenzo Servadei and
	Gianfranco Mauro and
	Thomas Stadelmayer and
	Avik Santra and
	Robert Wille},
	editor       = {M. Arif Wani and
	Mehmed M. Kantardzic and
	Vasile Palade and
	Daniel Neagu and
	Longzhi Yang and
	Kit Yan Chan},
	title        = {Uncertainty-based Meta-Reinforcement Learning for Robust Radar Tracking},
	booktitle    = {21st {IEEE} International Conference on Machine Learning and Applications,
	{ICMLA} 2022, Nassau, Bahamas, December 12-14, 2022},
	pages        = {1476--1483},
	publisher    = {{IEEE}},
	year         = {2022},
	doi          = {10.1109/ICMLA55696.2022.00232},
}

@ARTICLE{bgt_gesture_recogn,
	author={Hazra, Souvik and Santra, Avik},
	journal={IEEE Access}, 
	title={Short-Range Radar-Based Gesture Recognition System Using {3D} {CNN} With Triplet Loss}, 
	year={2019},
	volume={7},
	number={},
	pages={125623-125633},
	doi={10.1109/ACCESS.2019.2938725}}

@ARTICLE{bgt_gesture_recog2,
	author={Arsalan, Muhammad and Santra, Avik and Issakov, Vadim},
	journal={IEEE Transactions on Microwave Theory and Techniques}, 
	title={{RadarSNN}: A Resource Efficient Gesture Sensing System Based on mm-Wave Radar}, 
	year={2022},
	volume={70},
	number={4},
	pages={2451-2461},
	doi={10.1109/TMTT.2022.3148403}}

@INPROCEEDINGS{bgt_activ_recog,
	author={Khodabakhshandeh, Hamid and Visentin, Tristan and Hernangómez, Rodrigo and Pütz, Miro},
	booktitle={2021 21st International Radar Symposium (IRS)}, 
	title={Domain Adaptation Across Configurations of {FMCW} Radar for Deep Learning Based Human Activity Classification}, 
	year={2021},
	volume={},
	number={},
	pages={1-10},
	doi={10.23919/IRS51887.2021.9466179}}

@ARTICLE{bgt_activ_class,
	author={Stadelmayer, Thomas and Santra, Avik and Weigel, Robert and Lurz, Fabian},
	journal={IEEE Sensors Journal}, 
	title={Data-Driven Radar Processing Using a Parametric Convolutional Neural Network for Human Activity Classification}, 
	year={2021},
	volume={21},
	number={17},
	pages={19529-19540},
	doi={10.1109/JSEN.2021.3092002}}

@INPROCEEDINGS{bgt_indoor_loc,
	author={Thormann, Kolja and Steuernagel, Simon and Baum, Marcus},
	booktitle={2024 27th International Conference on Information Fusion (FUSION)}, 
	title={Indoor Localization based on Short-Range Radar and Rotating Landmarks}, 
	year={2024},
	volume={},
	number={},
	pages={1-8},
	doi={10.23919/FUSION59988.2024.10706316}}

@ARTICLE{music,
	author={Schmidt, R.},
	journal={IEEE Transactions on Antennas and Propagation}, 
	title={Multiple emitter location and signal parameter estimation}, 
	year={1986},
	volume={34},
	number={3},
	pages={276-280},
	doi={10.1109/TAP.1986.1143830}}

@ARTICLE{relax,
	author={Li, Jian and Stoica, P.},
	journal={IEEE Transactions on Signal Processing}, 
	title={Efficient mixed-spectrum estimation with applications to target feature extraction}, 
	year={1996},
	volume={44},
	number={2},
	pages={281-295},
	doi={10.1109/78.485924}}

@INPROCEEDINGS{sar_super_vit,
	author={Smith, Josiah W. and Alimam, Yusef and Vedula, Geetika and Torlak, Murat},
	booktitle={2022 IEEE Texas Symposium on Wireless and Microwave Circuits and Systems (WMCS)}, 
	title={A Vision Transformer Approach for Efficient Near-Field {SAR} Super-Resolution under Array Perturbation}, 
	year={2022},
	volume={},
	number={},
	pages={1-6},
	doi={10.1109/WMCS55582.2022.9866326}}

@INPROCEEDINGS{radar_swin_od,
	author={Giroux, James and Bouchard, Martin and Laganière, Robert},
	booktitle={2023 IEEE/CVF International Conference on Computer Vision Workshops (ICCVW)}, 
	title={{T-FFTRadNet}: Object Detection with {Swin} Vision Transformers from Raw {ADC} Radar Signals}, 
	year={2023},
	volume={},
	number={},
	pages={4032-4041},
	doi={10.1109/ICCVW60793.2023.00435}}

@ARTICLE{azimuth_superres_rDa,
	author={Schuessler, Christian and Hoffmann, Marcel and Vossiek, Martin},
	journal={IEEE Journal of Microwaves}, 
	title={Super-Resolution Radar Imaging With Sparse Arrays Using a Deep Neural Network Trained With Enhanced Virtual Data}, 
	year={2023},
	volume={3},
	number={3},
	pages={980-993},
	doi={10.1109/JMW.2023.3285610}}

@article{radar_swin_hrr,
	author = {Chen, Siyu and Huang, Xiaohong and Xu, Weibo},
	title = {Adaptive soft threshold transformer for radar high-resolution range profile target recognition},
	journal = {IET Radar, Sonar \& Navigation},
	volume = {18},
	number = {8},
	pages = {1260-1273},
	doi = {https://doi.org/10.1049/rsn2.12563},
	eprint = {https://ietresearch.onlinelibrary.wiley.com/doi/pdf/10.1049/rsn2.12563},
	year = {2024}
}

@Article{radar_super_sar2,
	AUTHOR = {Chen, Ruoxuan and Chen, Yumin and Zhang, Tengfei and Zeng, Fei and Li, Zhanghui},
	TITLE = {A Transformer-Based Residual Attention Network Combining {SAR} and Terrain Features for {DEM} Super-Resolution Reconstruction},
	JOURNAL = {Remote Sensing},
	VOLUME = {17},
	YEAR = {2025},
	NUMBER = {21},
	ARTICLE-NUMBER = {3625},
	ISSN = {2072-4292},
	DOI = {10.3390/rs17213625}
}

@misc{ada_frontend_audio,
	title={Adaptive Per-Channel Energy Normalization Front-end for Robust Audio Signal Processing}, 
	author={Hanyu Meng and Vidhyasaharan Sethu and Eliathamby Ambikairajah and Qiquan Zhang and Haizhou Li},
	year={2026},
	eprint={2510.18206},
	archivePrefix={arXiv},
	url={https://arxiv.org/abs/2510.18206}, 
	primaryClass={eess.AS},
}

@INPROCEEDINGS{trained_frontend_orig,
	author={Wang, Yuxuan and Getreuer, Pascal and Hughes, Thad and Lyon, Richard F. and Saurous, Rif A.},
	booktitle={2017 IEEE International Conference on Acoustics, Speech and Signal Processing (ICASSP)}, 
	title={Trainable frontend for robust and far-field keyword spotting}, 
	year={2017},
	volume={},
	number={},
	pages={5670-5674},
	doi={10.1109/ICASSP.2017.7953242}}

@inproceedings{radar_trans_indoor,
	author = {Yataka, Ryoma and Cardace, Adriano and Wang, Pu and Boufounos, Petros and Takahashi, Ryuhei},
	booktitle = {Advances in Neural Information Processing Systems},
	doi = {10.52202/079017-0625},
	editor = {A. Globerson and L. Mackey and D. Belgrave and A. Fan and U. Paquet and J. Tomczak and C. Zhang},
	pages = {19839--19869},
	publisher = {Curran Associates, Inc.},
	title = {{RETR}: Multi-View Radar Detection Transformer for Indoor Perception},
	volume = {37},
	year = {2024}
}

@ARTICLE{radar_trans_od_vit_prior,
	author={Cheng, Lei and Cao, Siyang},
	journal={IEEE Transactions on Radar Systems}, 
	title={TransRAD: Retentive Vision Transformer for Enhanced Radar Object Detection}, 
	year={2025},
	volume={3},
	number={},
	pages={303-317},
	doi={10.1109/TRS.2025.3537604}}

@misc{gelu,
	title={Gaussian Error Linear Units ({GELUs})}, 
	author={Dan Hendrycks and Kevin Gimpel},
	year={2023},
	eprint={1606.08415},
	archivePrefix={arXiv},
	primaryClass={cs.LG},
	url={https://arxiv.org/abs/1606.08415}, 
}

@ARTICLE{swin_1D_time_series,
	author={Chen, Guanyu and Shi, Tianyi and Xie, Baoxing and Zhao, Zhicheng and Meng, Zhu and Huang, Yadong and Dong, Jin},
	journal={IEEE Journal of Biomedical and Health Informatics}, 
	title={{SwinDAE}: Electrocardiogram Quality Assessment Using {1D} {Swin} Transformer and Denoising AutoEncoder}, 
	year={2023},
	volume={27},
	number={12},
	pages={5779-5790},
	doi={10.1109/JBHI.2023.3314698}}

@misc{swin_arxiv,
	title={Swin Transformer: Hierarchical Vision Transformer using Shifted Windows}, 
	author={Ze Liu and Yutong Lin and Yue Cao and Han Hu and Yixuan Wei and Zheng Zhang and Stephen Lin and Baining Guo},
	year={2021},
	eprint={2103.14030},
	archivePrefix={arXiv},
	primaryClass={cs.CV},
	url={https://arxiv.org/abs/2103.14030}, 
}

@InProceedings{stochastic_depth,
	author="Huang, Gao
	and Sun, Yu
	and Liu, Zhuang
	and Sedra, Daniel
	and Weinberger, Kilian Q.",
	editor="Leibe, Bastian
	and Matas, Jiri
	and Sebe, Nicu
	and Welling, Max",
	title="Deep Networks with Stochastic Depth",
	booktitle="Computer Vision -- ECCV 2016",
	year="2016",
	publisher="Springer International Publishing",
	address="Cham",
	pages="646--661",
	isbn="978-3-319-46493-0"
}

@inproceedings{adam,
	author       = {Diederik P. Kingma and
	Jimmy Ba},
	editor       = {Yoshua Bengio and
	Yann LeCun},
	title        = {Adam: {A} Method for Stochastic Optimization},
	booktitle    = {3rd International Conference on Learning Representations, {ICLR} 2015,
	San Diego, CA, USA, May 7-9, 2015, Conference Track Proceedings},
	year         = {2015},
	url          = {http://arxiv.org/abs/1412.6980},
	bibsource    = {dblp computer science bibliography, https://dblp.org}
}

@ARTICLE{lsd,
	author={Gray, A. and Markel, J.},
	journal={IEEE Transactions on Acoustics, Speech, and Signal Processing}, 
	title={Distance measures for speech processing}, 
	year={1976},
	volume={24},
	number={5},
	pages={380-391},
	doi={10.1109/TASSP.1976.1162849}}

@INPROCEEDINGS{audio_losses,
	author={Braun, Sebastian and Tashev, Ivan},
	booktitle={2021 44th International Conference on Telecommunications and Signal Processing (TSP)}, 
	title={A consolidated view of loss functions for supervised deep learning-based speech enhancement}, 
	year={2021},
	volume={},
	number={},
	pages={72-76},
	doi={10.1109/TSP52935.2021.9522648}}

@article{image_sr_review,
	title = {Image super-resolution: A comprehensive review, recent trends, challenges and applications},
	journal = {Information Fusion},
	volume = {91},
	pages = {230-260},
	year = {2023},
	issn = {1566-2535},
	author = {Dawa Chyophel Lepcha and Bhawna Goyal and Ayush Dogra and Vishal Goyal},
}

@INPROCEEDINGS{ResNet,
	author={He, Kaiming and Zhang, Xiangyu and Ren, Shaoqing and Sun, Jian},
	booktitle={2016 IEEE Conference on Computer Vision and Pattern Recognition (CVPR)}, 
	title={Deep Residual Learning for Image Recognition}, 
	year={2016},
	volume={},
	number={},
	pages={770-778},
	doi={10.1109/CVPR.2016.90}}

@misc{radar_foundation,
	title={Towards Foundational Models for Single-Chip Radar}, 
	author={Tianshu Huang and Akarsh Prabhakara and Chuhan Chen and Jay Karhade and Deva Ramanan and Matthew O'Toole and Anthony Rowe},
	year={2025},
	eprint={2509.12482},
	archivePrefix={arXiv},
	primaryClass={cs.CV},
	url={https://arxiv.org/abs/2509.12482}, 
}

@InProceedings{detr,
	author="Carion, Nicolas
	and Massa, Francisco
	and Synnaeve, Gabriel
	and Usunier, Nicolas
	and Kirillov, Alexander
	and Zagoruyko, Sergey",
	editor="Vedaldi, Andrea
	and Bischof, Horst
	and Brox, Thomas
	and Frahm, Jan-Michael",
	title="End-to-End Object Detection with Transformers",
	booktitle="Computer Vision -- ECCV 2020",
	year="2020",
	publisher="Springer International Publishing",
	address="Cham",
	pages="213--229",
	isbn="978-3-030-58452-8"
}

@ARTICLE{sar_superres_cvcnn,
	author={Smith, Josiah W. and Torlak, Murat},
	journal={IEEE Transactions on Aerospace and Electronic Systems}, 
	title={Deep-Learning-Based Multiband Signal Fusion for 3-{D} {SAR} Superresolution}, 
	year={2024},
	volume={60},
	number={1},
	pages={8-24},
	doi={10.1109/TAES.2023.3270111}}

@ARTICLE{cvnn_hrr,
	author={Pan, Pingping and Zhang, Yunjian and Deng, Zhenmiao and Wu, Gang},
	journal={IEEE Transactions on Geoscience and Remote Sensing}, 
	title={Complex-Valued Frequency Estimation Network and Its Applications to Superresolution of Radar Range Profiles}, 
	year={2022},
	volume={60},
	number={},
	pages={1-12},
	doi={10.1109/TGRS.2021.3117298}}

@ARTICLE{swin_1D_hrr,
	author={Smith, Josiah W. and Torlak, Murat},
	journal={IEEE Geoscience and Remote Sensing Letters}, 
	title={Frequency Estimation Using Complex-Valued Shifted Window Transformer}, 
	year={2024},
	volume={21},
	number={},
	pages={1-5},
	doi={10.1109/LGRS.2024.3411554}}

@misc{cfar_loss_diffusion_augmentation,
	title={Synthetic {FMCW} Radar Range Azimuth Maps Augmentation with Generative Diffusion Model}, 
	author={Zhaoze Wang and Changxu Zhang and Tai Fei and Christopher Grimm and Yi Jin and Claas Tebruegge and Ernst Warsitz and Markus Gardill},
	year={2026},
	eprint={2601.06228},
	archivePrefix={arXiv},
	primaryClass={cs.CV},
	url={https://arxiv.org/abs/2601.06228}, 
}

@INPROCEEDINGS{pix2pix,
	author={Isola, Phillip and Zhu, Jun-Yan and Zhou, Tinghui and Efros, Alexei A.},
	booktitle={2017 IEEE Conference on Computer Vision and Pattern Recognition (CVPR)}, 
	title={Image-to-Image Translation with Conditional Adversarial Networks}, 
	year={2017},
	volume={},
	number={},
	pages={5967-5976},
	doi={10.1109/CVPR.2017.632}}

@INPROCEEDINGS{perc_dist_tradeoff,
	author={Blau, Yochai and Michaeli, Tomer},
	booktitle={2018 IEEE/CVF Conference on Computer Vision and Pattern Recognition}, 
	title={The Perception-Distortion Tradeoff}, 
	year={2018},
	volume={},
	number={},
	pages={6228-6237},
	doi={10.1109/CVPR.2018.00652}}

@InProceedings{dmatching_hallu,
	author="Cohen, Joseph Paul
	and Luck, Margaux
	and Honari, Sina",
	editor="Frangi, Alejandro F.
	and Schnabel, Julia A.
	and Davatzikos, Christos
	and Alberola-L{\'o}pez, Carlos
	and Fichtinger, Gabor",
	title="Distribution Matching Losses Can Hallucinate Features in Medical Image Translation",
	booktitle="Medical Image Computing and Computer Assisted Intervention -- MICCAI 2018",
	year="2018",
	publisher="Springer International Publishing",
	address="Cham",
	pages="529--536",
	isbn="978-3-030-00928-1"
}

@INPROCEEDINGS{SRGAN,
	author={Ledig, Christian and Theis, Lucas and Huszár, Ferenc and Caballero, Jose and Cunningham, Andrew and Acosta, Alejandro and Aitken, Andrew and Tejani, Alykhan and Totz, Johannes and Wang, Zehan and Shi, Wenzhe},
	booktitle={2017 IEEE Conference on Computer Vision and Pattern Recognition (CVPR)}, 
	title={Photo-Realistic Single Image Super-Resolution Using a Generative Adversarial Network}, 
	year={2017},
	volume={},
	number={},
	pages={105-114},
	doi={10.1109/CVPR.2017.19}}

\vspace{11pt}
\begin{IEEEbiography}[{\includegraphics[width=1in,height=1.25in,clip,keepaspectratio]{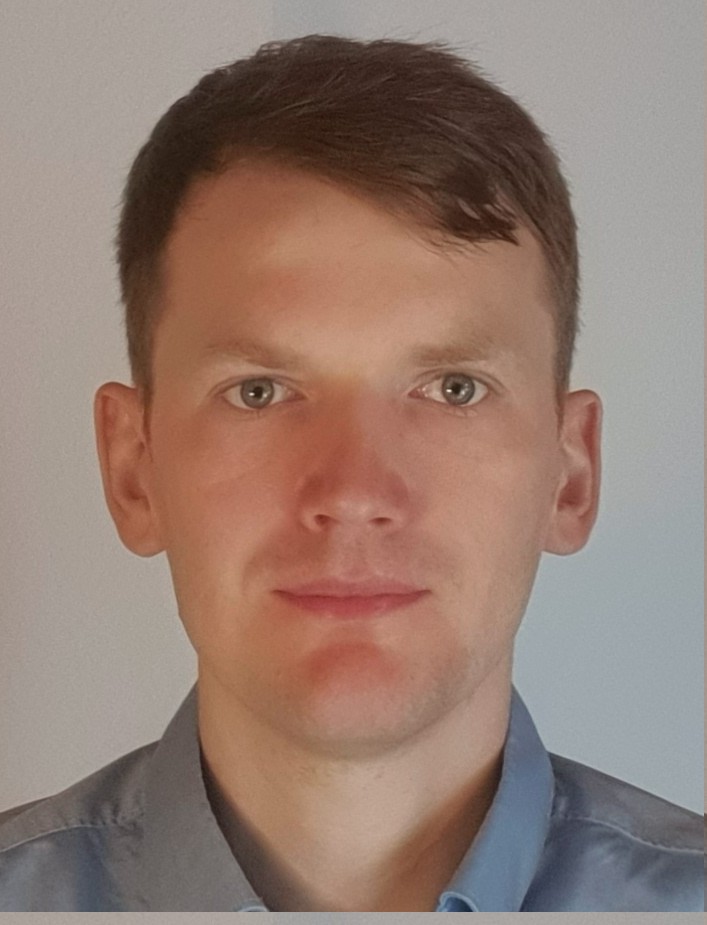}}]{Sven Hinderer}
received the B.Sc. and M.Sc. in electrical engineering \& information technology from the University of Stuttgart in 2015 and 2018, respectively. From 2019 to 2022 he was with Fraunhofer FHR, Wachtberg, Germany, studying the application of deep learning in passive surveillance. Since 2022, he's with the Institute of Signal Processing and System Theory, University of Stuttgart, Germany, pursuing his PhD in low-cost indoor localization. His research interests include localization \& tracking, machine learning, radar, audio, and computer vision.
\end{IEEEbiography}
\begin{IEEEbiography}[{\includegraphics[width=1in,height=1.25in,clip,keepaspectratio]{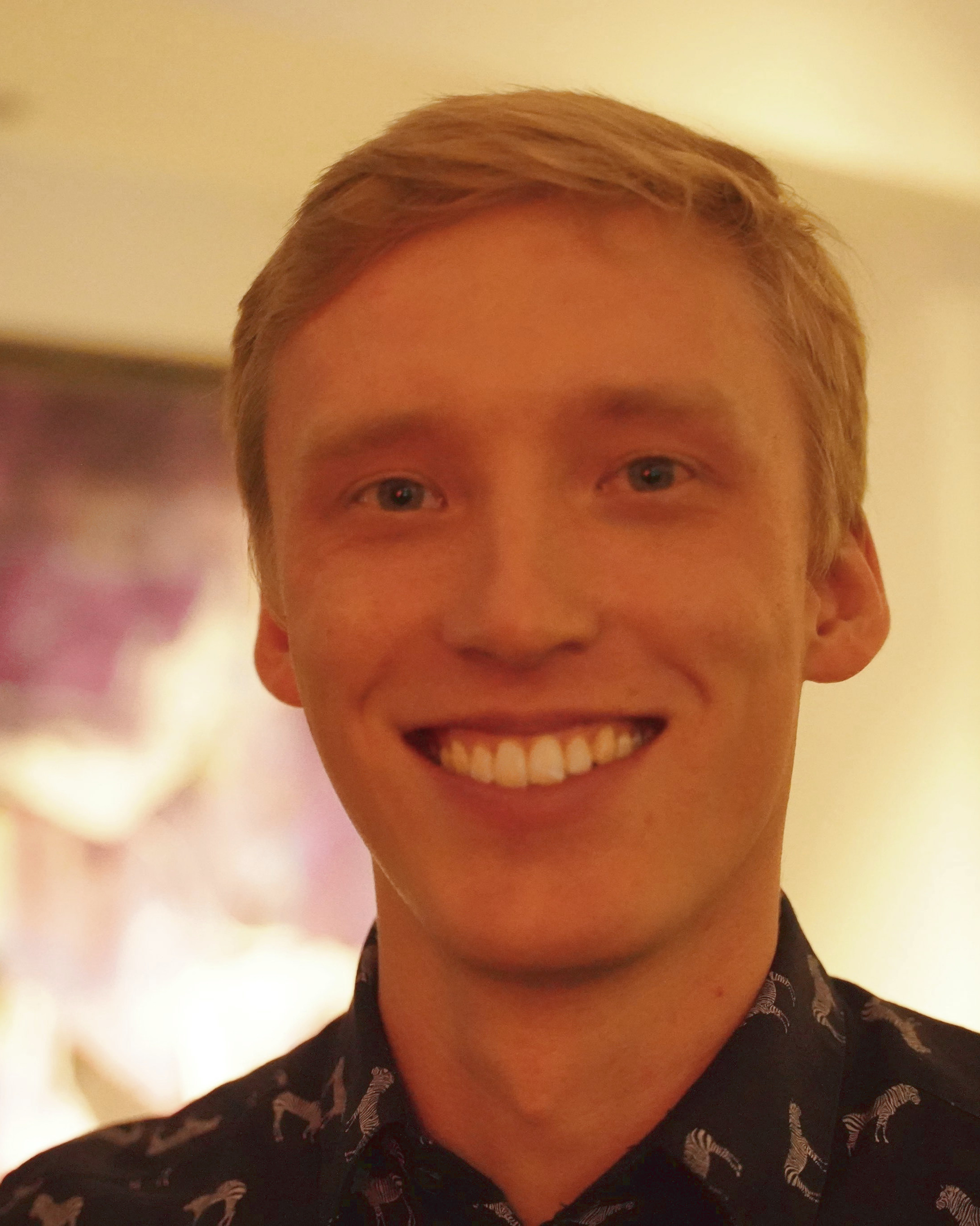}}]{Jonathan Riese}
	received his B.Sc. degree in vehicle and engine engineering from the University of Stuttgart in 2020 and is currently pursuing his M.Sc. degree in mechatronics there.
	His research interests include machine learning, signal processing and control theory \& engineering, especially model predictive control.
\end{IEEEbiography}
\begin{IEEEbiography}[{\includegraphics[width=1in,height=1.25in,clip,keepaspectratio]{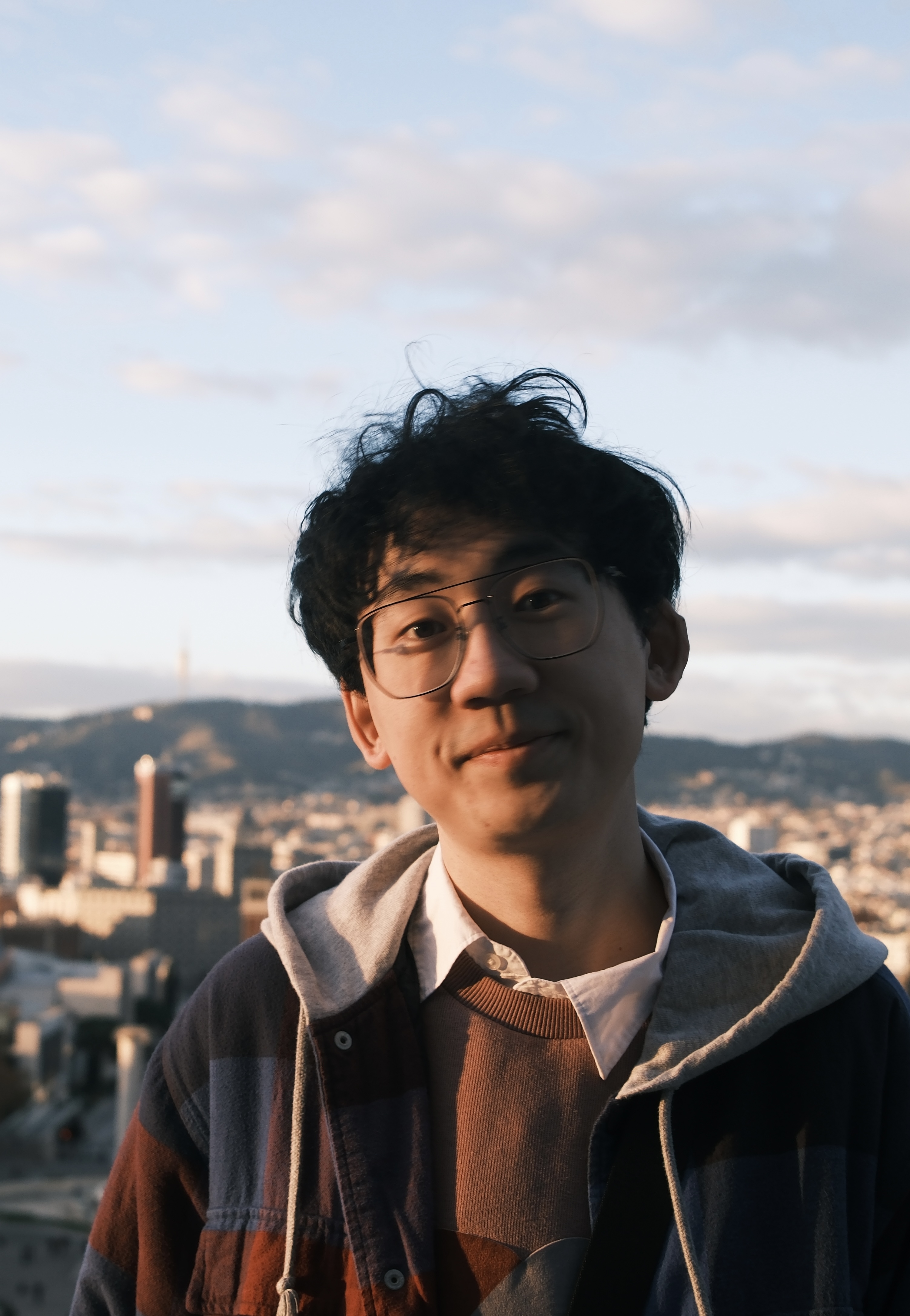}}]{Zheming Yin}
	received the B.Eng. degree in vehicle engineering from Northeastern University, China, in 2021, and the M.Sc. degree in electromobility from the University of Stuttgart, Germany, in 2025. His research interests are interdisciplinary, focusing on signal processing, human-computer interaction, autonomous driving, and machine learning.
\end{IEEEbiography}
\begin{IEEEbiography}[{\includegraphics[width=1in,height=1.25in,clip,keepaspectratio]{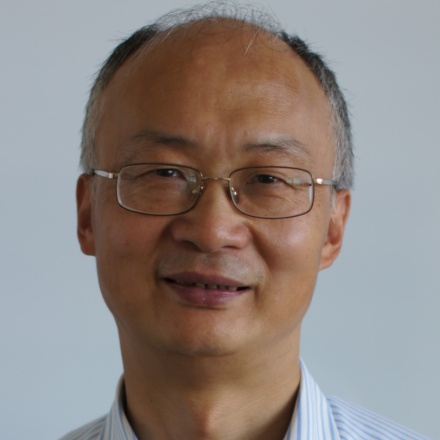}}]{Bin Yang}(Senior Member, IEEE) received the
	Dipl.-Ing. and Ph.D. degrees in electrical engineering in Germany. He is a Full Professor and
	the head of the Institute of Signal Processing and System Theory, University of Stuttgart,
	Germany. His research interests include methods
	and algorithms of statistical signal processing,
	machine learning, and in particular deep learning.
	The research spectrum covers discriminative models, generative models, self-supervised learning,
	domain adaptation, anomaly detection, and causal reasoning, in connection
	with a variety of applications like medical imaging, autonomous driving,
	radar, speech, localization, and semiconductor test.
\end{IEEEbiography}
\end{document}